\documentclass[useAMS,usenatbib]{mn2e}
\usepackage{graphicx}
\usepackage{amssymb}
\usepackage{lscape}
\usepackage{rotating}
\usepackage{subfigure}

\def\deg{\ifmmode^\circ\else$^\circ$\fi}

\def\Q{\ifmmode\mathcal{Q}\else$\mathcal{Q}$\fi}
\def\Mach{\ifmmode\mathcal{M}\else$\mathcal{M}$\fi}

\title[Infrared study of S235 complex]
{Infrared photometric study of the massive star forming region S235 using {\it Spitzer}-IRAC and JHK observations}
\author[L.K. Dewangan \& B.G. Anandarao]
{L.K. Dewangan$^{1}$\thanks{lokeshd@prl.res.in}, \& B.G. Anandarao$^{1}$\thanks{anand@prl.res.in}\\
$^1$Astronomy $\&$ Astrophysics Division, Physical Research Laboratory, Navrangpura, Ahmedabad 380 009, India.\\}

\begin{document}

\date{ }

\pagerange{\pageref{firstpage}--\pageref{lastpage}} \pubyear{2011}

\maketitle

\label{firstpage}

\begin{abstract}
We present the {\it Spitzer}-IRAC images of the S235 star forming complex that
includes the East~1 \& 2, Central and S235 A \& B regions. In addition, we present the
near-infrared images of the S235 A \& B regions. The IRAC photometry reveals 
on-going star formation, with 86 
Class 0/I and 144 Class II YSOs in the entire S235 complex. Nearly 73\% of 
these YSOs are present in clusters with a maximum surface density of 120 
YSOs/pc$^{2}$ (in the vicinity of S235A \& B regions). A few YSOs, possibly in an arc-like  
formation, are identified towards the south of S235A region, which may be speculated as an 
evidence for magnetically super-critical collapse. One of the sources in 
the arc-like formation, namely S235AB-MIR, seems to be a young, massive star 
that is still accreting matter. SED modeling of some of the newly identified YSOs confirms the
classification made on the basis of IRAC colours. The IRAC ratio map of Ch2/Ch4 traces clearly the 
Br$\alpha$ emission associated with the HII region of S235A within the horse-shoe 
envelope. Outside the horse-shoe structure, the ratio map indicates 
shock-excited H$_{2}$ emission. Br$\alpha$ emission is also seen around
S235B (from the ratio map).  
The ratio map of Ch2/Ch4 reveals that the source ``e2s3" in the East~2 region may be 
associated with shock-excited H$_2$ emission outflow or jet. The SED modeling of this new 
source indicates that it is a very young massive star 
that is not yet able to drive an HII region.
\end{abstract}

\begin{keywords}
stars: formation -- stars: pre-main-sequence --  stars: winds and outflows 
-- infrared: ISM -- ISM: HII Regions -- ISM: Individual: S235
\end{keywords}

\section{Introduction}
\label{sec:intro}
The S235 is a large extended HII region associated with a star forming complex and situated at a distance 
between 1.6 and 2.5 
kpc \citep{georgelin73,israel78}, in the Perseus Spiral Arm. 
\citet{georgelin73} found that the S235 region is excited by the massive star BD +35$^{o}$1201 
of O9.5V type that has ionised and 
dispersed the surrounding molecular gas. The earlier studies on this complex show that 
it is a site of active star 
formation \citep{israel78,felli97,felli04,felli06,allen05,kirsanova08}. This region covers 
some well known star forming sites called 
East~1, East~2, Central \citep{kirsanova08} and S235A, S235B and S235C \citep{felli97}. \\\\
Using molecular gas kinematics and density distribution from $^{13}$CO(1-0) and CS(2-1) emission 
observations in the mm region, 
\citet{kirsanova08} presented evidences to argue that the expanding HII region in S235 triggered fresh
star formation in the north-east and north-west regions of the complex (``East~1, East~2, Central and North-West").
\citet{kirsanova08} argued that the star formation may have been triggered either by the ``collect and collapse" process or 
by shocking the already clumpy regions and that the denser East~1 cluster is probably younger 
than the Central and East~2 clusters. 
\citet{allen05} presented {\it Spitzer}-IRAC photometry on the S235 complex and \citet{felli06} studied some selected 
sources in the S235A and S235B regions (henceforth called as S235A-B) and identified several young protostars.
\citet{felli04} studied the region S235A-B with high-resolution mm lines (from the molecules HCO$^{+}$, C$^{34}$S, H$_{2}$CS, SO$_{2}$, and 
CH$_{3}$CN) and continuum observations, together with far-IR observations and reported two molecular outflows in HCO$^{+}$(1-0)
centred on the compact molecular core (traced by the 1.2 mm continuum observations).
Later on, from the VLA radio observations along with the {\it Spitzer}-IRAC observations on the S235A-B region, \citet{felli06} argued that the 
expanding HII region from the massive star in S235A could have triggered the star formation in the dense cluster between the S235A and S235B regions.
They also found a new embedded source S235AB-MIR in three IRAC bands (except in Ch1) in 
the mm core. Further, these authors detected two compact radio-sources, VLA-1 and VLA-2, that coincide with the source M1 and the 
centre of S235B respectively. \citet{Saito07} identified dense clumps from their observations of C$^{18}$O emission (that traces dense gas). 
\citet{krassner82} found recombination lines of Hydrogen and the PAH emission features at 3.3, 8.7 
and 11.3 $\mu$m associated with S235B. 
\citet{boley09} attributed the central star of S235B to a rare class of early type Herbig Be star.\\\\
In the backdrop of the existing observations and interpretations of the S235 complex, our aims to revisit the {\it Spitzer}-IRAC 
archival data are: (i) to find YSOs (systematically taking care of all possible contaminants), (ii) to estimate YSO surface 
densities, (iii) to identify some interesting YSOs and derive their physical parameters using SED modeling 
and (iv) to delineate regions of different emission lines/features (Br$\alpha$, H$_{2}$ and PAH) from IRAC ratio images. 
To our knowledge, these were not done before. In addition, we present new NIR photometry, using which in 
conjunction with IRAC photometry, we attempt to extract more YSOs in the vicinity of S235A-B.\\\\
In Section~\ref{sec:obser}, we describe the data used for the present study and the analysis tasks utilised. 
Section~\ref{sec:disc} presents the results and discussion on infrared photometry of embedded sources associated with the S235 complex. 
In this section, we also present the results and discussion on the ratio maps. In Section~\ref{sec:conc}, we give the conclusions.
\section{Observations and Data Analysis}
\label{sec:obser}
{\it Spitzer}-IRAC archival data around the source S235 (encompassing IRAS 05375+3540) were obtained 
from the Spitzer Science Center (SSC) on 6 May 2009. The observations for S235 were taken by {\it Spitzer} in the High Dynamic Range (HDR) mode 
with 12s integration time in all four filters on 12 March 2004 as a part of the GTO program (Id: 201; Project title:
``The Role of Photodissociation Regions in High Mass Star Formation"; PI: G.~Fazio). 
Basic Calibrated Data (BCDs) images were processed (using the pipeline version S14.0.0)  
for `jailbar' removal, saturation 
and `muxbleed' correction before making the final mosaic using Mopex and IDL softwares \citep{Makovoz05}. 
A pixel ratio (defined as the ratio of the area formed by the original pixel scale, 1.22 arcsec/px, to that of the 
mosaiced pixel scale) of 2 was adopted for making the mosaic (giving an
effective platescale of 0.86 arcsec/pixel). 
A total number of 72 BCD images of size 5.2 $\times$ 5.2 arcmin$^{2}$ were mosaiced to obtain a final image 
of size 24.6 $\times$ 22.0 arcmin$^{2}$ commonly in all the four bands. 
Aperture photometry was performed on the mosaiced images with 2.8 pixel aperture and sky annuli 
of 2.8 and 8.5 pixels using APPHOT task in the IRAF package. The zero points for these apertures (including 
aperture corrections) are, 17.80, 17.30, 16.70 and 15.88 mag for the 3.6, 4.5, 5.8, 8.0 $\mu$m bands 
(here onwards called as Ch1, Ch2, Ch3 and Ch4 respectively). The photometric uncertainties were found to vary 
between 0.01 to 0.25 for the four channels, with those for longer wavelengths (i.e., 5.8 and 8.0 $\mu$m) being 
on the higher side. We have considered sources common to all the four bands which have photometric uncertainties 
$\leq$ 0.20 in Ch1-3 and $\leq$ 0.25 in Ch4. The source matching in the bands 
was done by considering positional coincidence within 
a circle defined by the platescale ($\sim$ 1 arcsec). Following the procedure given by \citet{Qiu08}, 
we have estimated the completeness limit in Ch1 to be 15 mag with the completeness of about 78\%, 
considering the fact that the Ch1 of IRAC detects point sources fainter than the other bands.\\\\
Near-infrared (NIR) photometric observations were made on the massive star forming region 
using the Physical Research Laboratory (PRL) 1.2~m telescope at Mt. Abu, India. 
A region of 8$\times$8 arcmin$^{2}$ around the 
source IRAS 05375+3540 (in the vicinity of S235A and S235B regions in the S235 complex) was observed during 20-21 January 2010 under 
photometric conditions with an average seeing of 1.5-2 arcsec, 
using the recently commissioned Near-Infrared Camera and Spectrograph (NICS: \citet{anandarao08}) that has a 
1024 $\times$ 1024 HgCdTe array Wide-area Infrared Imager-I (HAWAII-1; Teledyne, USA) 
in the J(1.17-1.33 $\mu$m), H(1.49-1.78 $\mu$m) and K(2.03-2.37 $\mu$m) bands (new MKO filters: see \citet{toku02}), 
with a plate scale of 0.5 arcsec/pixel. Photometric calibration was done using 
the standard star AS11 \citep{hunt98}.
Limiting magnitudes of 17, 16.5 and 16.0 mag were achieved in integration times of 300, 200 and 200 seconds in J, H and K bands 
respectively. These limiting magnitudes are deeper than the 2MASS survey for this object. 
All the Mt Abu images were processed using standard pipeline procedures in IRAF software \citep{tody93}, 
like dark and sky subtraction and flat-fielding. 
The images were then co-added and averaged to obtain a final image in each band.
PSF photometry is performed on the images using DAOPHOT task in IRAF \citep{stetson87} 
to obtain individual point source magnitudes (details of the procedures of PSF photometry are given in \citet{longmore09}).
Common sources between the JHK and IRAC data are obtained by positional matching within a circle of 1 arcsec.

\section{Results and Discussion} 
\label{sec:disc}
Our work is mainly focused on the three regions, viz. S235A-B, East~1 and East~2 in the S235 molecular cloud complex. 
We have divided this section into two parts. In the first part, we discuss the IRAC photometry of the 
sources in the entire region, in general and of the three regions, in particular;  
we then describe the modeling of important and interesting 
sources and their possible formation scenario. In the second part, we describe the ratio 
maps produced from the IRAC images for the entire complex and discuss possible interpretations.
\subsection{IRAC Photometry}
\label{subsec:phot}
\begin{figure*}
\includegraphics[width=\textwidth]{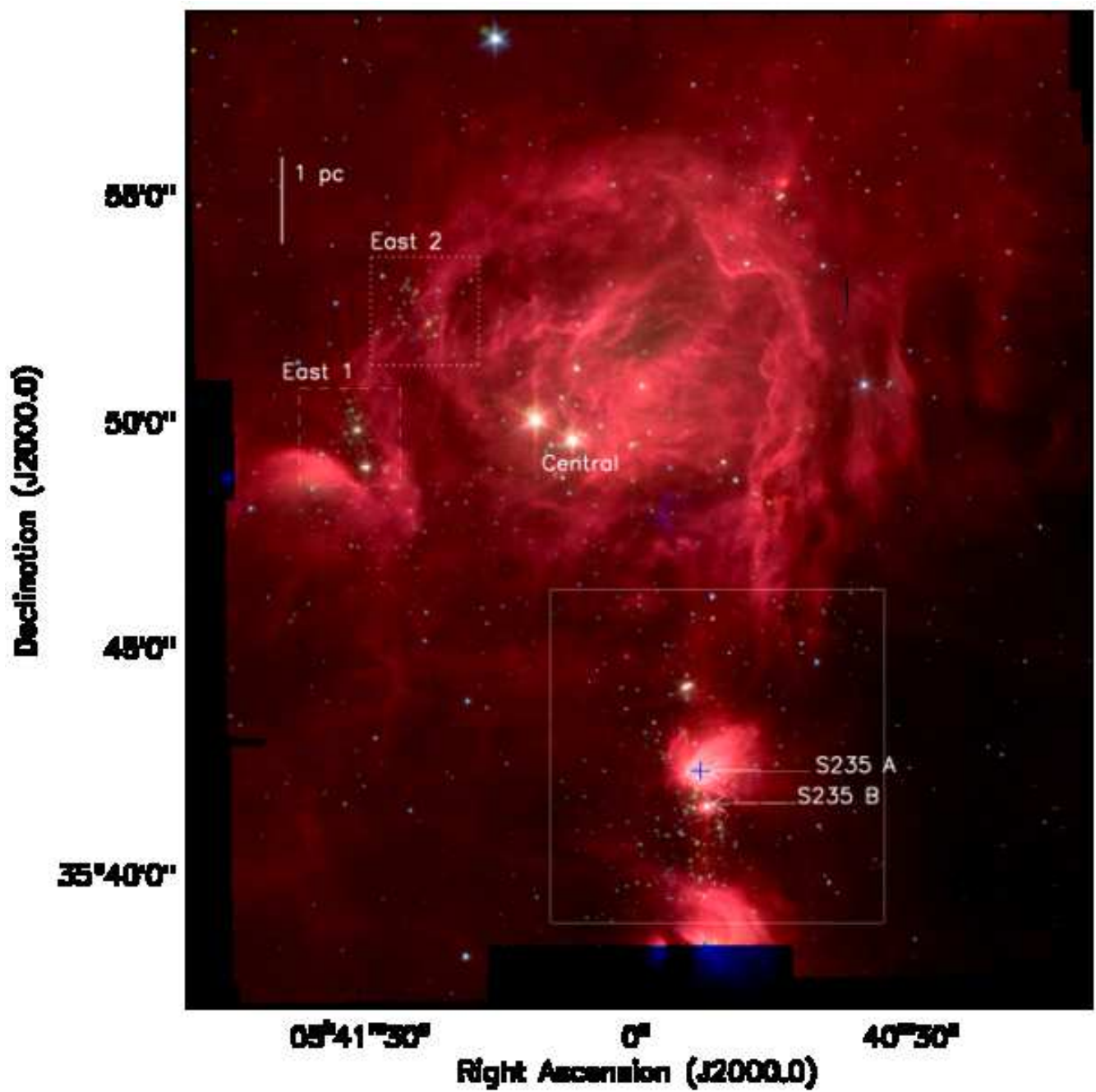}
\caption{IRAC 3-colour-composite image (size $\sim$ 24.6 $\times$ 22.0 arcmin$^{2}$) of S235 complex  
(8.0 (red), 4.5 (green) and 3.6 $\mu$m (blue)). Different regions identified by \citet{kirsanova08} are labeled in the 
region within S235 complex. Two of such regions in the complex, East~1 and East~2, are shown by dashed and dotted boxes respectively. 
The solid line box in the complex represents the region covered by Mt Abu JHK observations. The blue `plus' sign indicates the 
position of IRAS 05375+3540.}
\label{fig1}
\end{figure*}

\begin{figure*}
\includegraphics[width=0.58\textwidth]{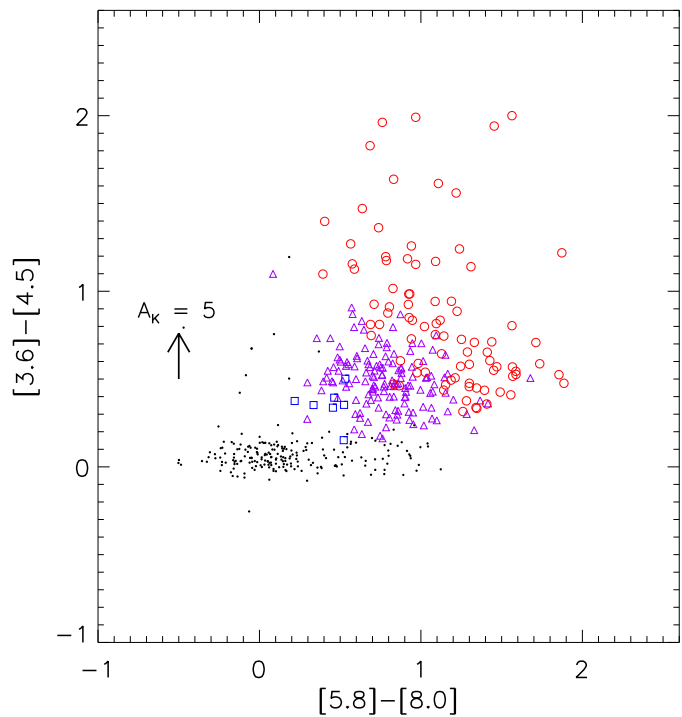}
\caption{Mid-IR colour-colour diagram using the {\it Spitzer}-IRAC bands for all the sources identified 
within the region shown in Fig.~\ref{fig1}. The extinction vector for A$_{K}$ = 5 mag is shown by the arrow on the left side, 
using average extinction law from \citet{Flaherty07}. The black dots around the centre (0,0) locate the stars with only photospheric emissions. 
The open squares (blue), open triangles (violet) and open circles (red) 
represent respectively, Class III, Class II and Class 0/I sources, classified by using the $\alpha_{IRAC}$ criteria.}
\label{fig2}
\end{figure*}

\begin{figure*}
\includegraphics[width=1.\textwidth]{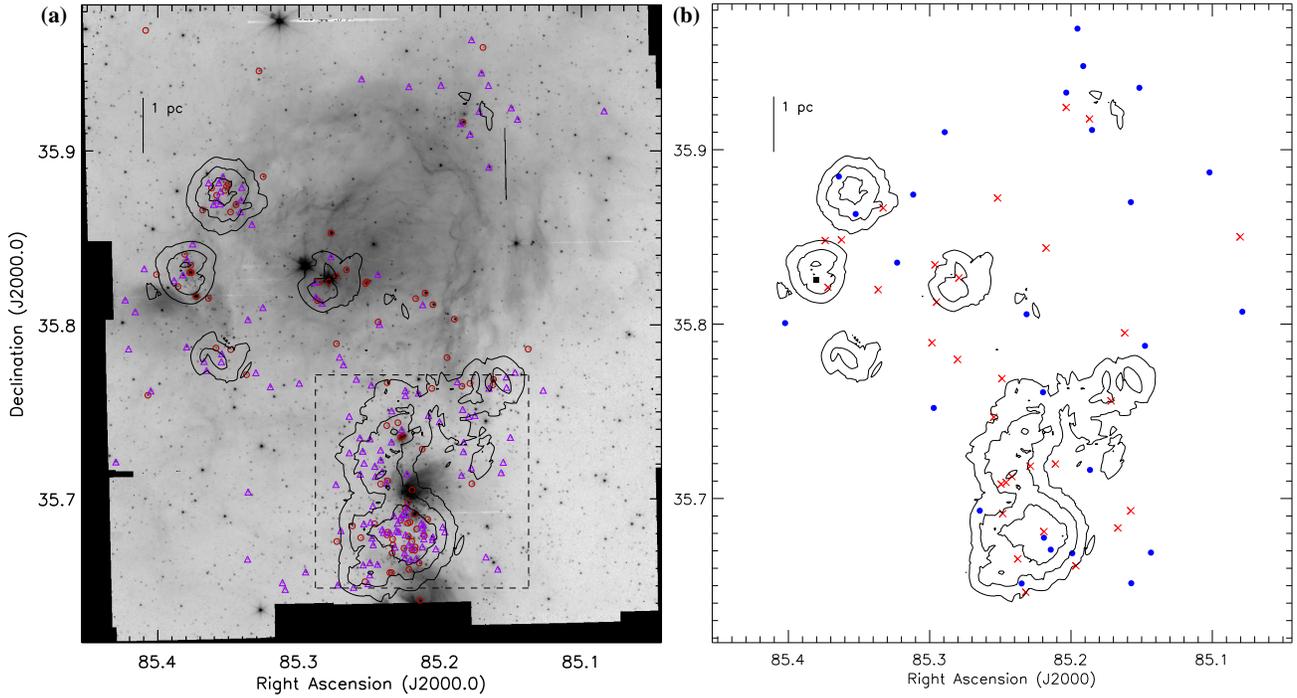}
\caption{a: Inverted {\it Spitzer}-IRAC Ch2 (4.5 $\mu$m) image of S235 (similar size as shown 
in Fig.~\ref{fig1}) over-plotted by IRAC Class 0/I and II sources. The open circles (red) 
and open triangles (violet) show the IRAC Class 0/I and II sources respectively. Dashed box represents the NIR observed 
region as shown in Fig.~\ref{fig1} by a solid box.
b: S235 region showing contaminants: crosses (red) show resolved PAH aperture 
contaminants, filled squares (black) show unresolved shocked emission knots and the 
filled circle (blue) the PAH galaxy contaminant.
In both the figures, the contours show YSO iso-density at 5, 10 and 30 YSOs/pc$^{2}$ levels, from outer to inner side
(see text for details).} 
\label{fig3}
\end{figure*}

In this sub-section, we present a photometric study carried out on the whole S235 complex using the four IRAC bands.
A 3-colour-composite image of the IRAC bands (Ch4(red), Ch2(green) and Ch1(blue)) is shown in Fig.~\ref{fig1}, encompassing the East~1, East~2, Central and 
S235A-B regions (marked in the Fig).\\\\
IRAC colour-colour and colour-magnitude spaces are a very powerful tool to identify the deeply embedded YSOs in the 
star forming regions. One problem with this is, however, that there is a possibility of contamination by non-YSO 
candidates like PAH galaxies, broad-line AGNs, unresolved shocked emission blobs/knots, PAH-emission contaminated 
apertures and Asymptotic Giant Branch (AGB) stars whose colours closely resemble those of genuine YSOs.
Therefore, it is important to identify the correct YSOs with the removal of these contaminants from the sources 
identified in the star forming regions. We used the criteria given by \citet{Gutermuth09} to separate the YSOs 
from the contaminants. 
We have obtained a total of 507 point sources that are commonly identified in all the four IRAC bands. Of these 507 
sources, 58 are found to be contaminants (26 PAH galaxies, 1 shocked emission and 
31 PAH aperture-contaminants), 
while there are 237 YSOs and 212 purely photospheric sources.\\\\
After the identification and removal of the contaminants, we used the criteria based on the 
spectral index ($\alpha_{IRAC}$), to classify the YSOs into different evolutionary classes (see e.g., \citet{Green94,Smith04,Lada06}).
Using the specific criteria for IRAC bands given in \citet{Billot10} and the references therein, 
we obtained 86 Class 0/I, 144 Class II sources 
(a total of 230 YSOs) and 7 Class III sources. It may be noted here that the so-called flat-spectrum 
sources with in-falling envelopes are added to the Class 0/I list (see \citet{Billot10}). Such sources 
are actually border-line cases between Class I and Class II. 
The locations of the identified YSOs (Class 0/I and II), Class III as well as the 
photospheric sources are plotted in the IRAC colour-colour diagram shown in Fig.~\ref{fig2}. 
The sky coordinates, IRAC magnitudes and spectral indices of Class 0/I and Class II sources are given in 
Tables~\ref{tab1} and \ref{tab2} respectively. The tables also contain the JHK magnitudes with their uncertainties: 
the ones in italics are from Mt Abu observations (MKO system) and the rest are 
from 2MASS archives (having S/N $\geq$ 5 or flag C).\\\\
In order to estimate the clustering of YSOs in the S235 complex, we have calculated the un-biased surface 
density of YSOs using a 5 arcsec grid size, following the same procedure as given in \citet{Gutermuth09} and the references
therein. 
The surface density of YSOs is estimated using 6 nearest-neighbour (NN) YSOs for each grid point. 
Fig.~\ref{fig3}a shows the spatial distribution of YSOs (Class 0/I sources in red open circles and 
Class II in violet open triangles) overlaid on the IRAC Ch2 image (4.5 $\mu$m). The YSO surface density 
contours are also shown in the Figure, with contour levels of 30, 10 and 5 YSOs/pc$^{2}$ decreasing from inner to outer side.
Fig.~\ref{fig3}b shows the distribution of contaminants.  
It may be noticed that the distribution of YSOs is mostly concentrated in the East~1, East~2 and the vicinity of 
S235A-B region, while YSO densities of $\sim$ 10 YSOs/pc$^{2}$ are also seen around the Central region and south of East 1. 
The maximum densities 
are about 50 YSOs/pc$^{2}$ in the complex. These densities are comparable to those obtained for other 
star forming regions reported elsewhere (e.g., \citet{Chavarria08, Billot10, lokesh10}). 
In addition, we have calculated the empirical cumulative distribution (ECD) as a function of NN distance  
to identify the clustered YSOs. 
Using the ECD, we have estimated the distance of inflection ($d_{c}$ that represents 
the maximum separation between the cluster members) to identify 
the cluster members within the contour level 5 YSOs/pc$^{2}$ in the entire complex of S235 as well as in the 
individual smaller regions. We obtain $d_{c}$ = 0.71 pc in the entire complex.  
We find that 73\% of 
the YSOs are present in clusters concentrated in the regions, East~1, East~2, Central, 
south of East 1 and the vicinity of S235A-B. 
The locations of maxima of YSO surface density are consistent with the dense regions traced out by CS(2-1) emission.
We find the highest density of YSOs in the S235A-B region among all the clusters in the entire S235 complex.\\\\
The possibility of ``red-source" contamination among the scattered YSOs is discussed in Appendix A.

%
%
\subsubsection{S235-East~1} 
\begin{figure*}
\includegraphics[width=0.60\textwidth]{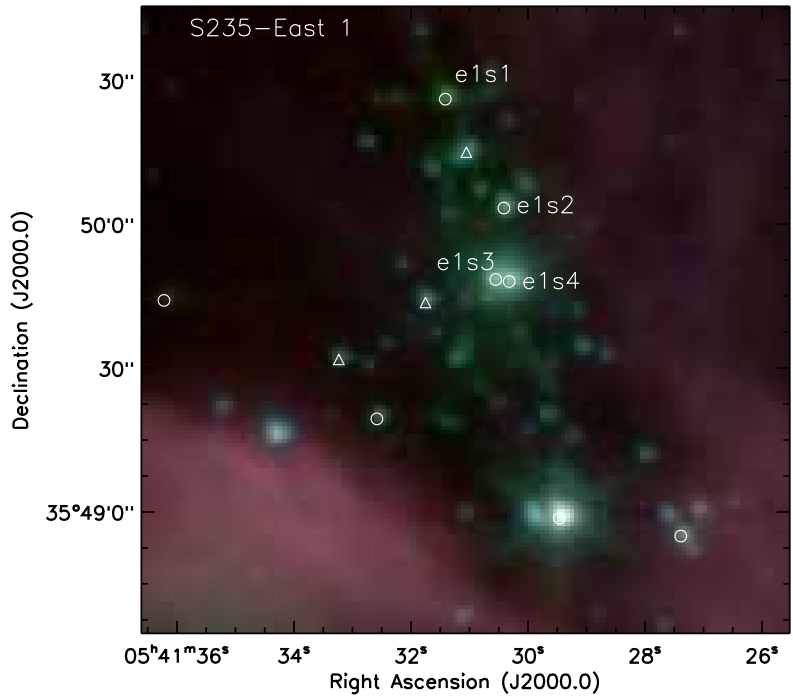}
\caption{IRAC 3-colour-composite (Ch4(red), Ch2(green) and Ch1(blue)) zoomed-in image of East~1 region (cf. Fig.~\ref{fig1}: dashed box).
Classified YSOs (Class 0/I in open circles and Class II in open triangles) are overlaid on the image.
The labeled sources are Class 0/I YSOs that are not detected in 2MASS JHK image. 
The image shows that the green colour (i.e., IRAC Ch2 (4.5 $\mu$m)) is more dominant around the marked sources. 
This may be associated with the shock-excited molecular H$_{2}$ emission because IRAC 
Ch2 (4.5 $\mu$m) is sensitive for shock-excited molecular H$_{2}$ emission.}
\label{fig4}
\end{figure*}

The East~1 region is shown magnified in Fig.~\ref{fig4} using IRAC 3-colour-composite image from 
Fig.~\ref{fig1} (dashed box).
We found 8 Class 0/I and 3 Class II sources in the East~1 region (marked in the Figure by open circles and triangles respectively).
Some of the Class 0/I sources that do not have JHK counterparts are labeled as `e1s1-e1s4' in
Fig.~\ref{fig4}.
The labeled sources are given in Table~\ref{tab3} with their sky positions and observed IRAC magnitudes.
The distance of inflection for this cluster is $d_{c}$ = 0.52 pc.
The IRAC colour-composite image shows the presence of the shock-excited molecular H$_{2}$ emission in the East~1 region,
as may be inferred by the predominance of Ch2 band (in green colour).
We did the SED modeling using an on-line tool \citep{Robit06,Robit07} for all the labeled sources in S235 East~1 region.
Only those models are selected that satisfy the 
criterion $\chi^{2}_{best}$ - $\chi^{2}$ $<$ 3, where $\chi^{2}$ is taken per data point.
The weighted mean values of the physical parameters derived from SED modeling for all selected sources
are given in Table~\ref{tab4}. 
The SED model results show that all the four labeled sources are young and still going through the accretion phase.
These SED results are also discussed in the light of the IRAC ratio maps (see Sec. 3.2).
\subsubsection{S235-East~2}
\begin{figure*}
\includegraphics[width=0.60\textwidth]{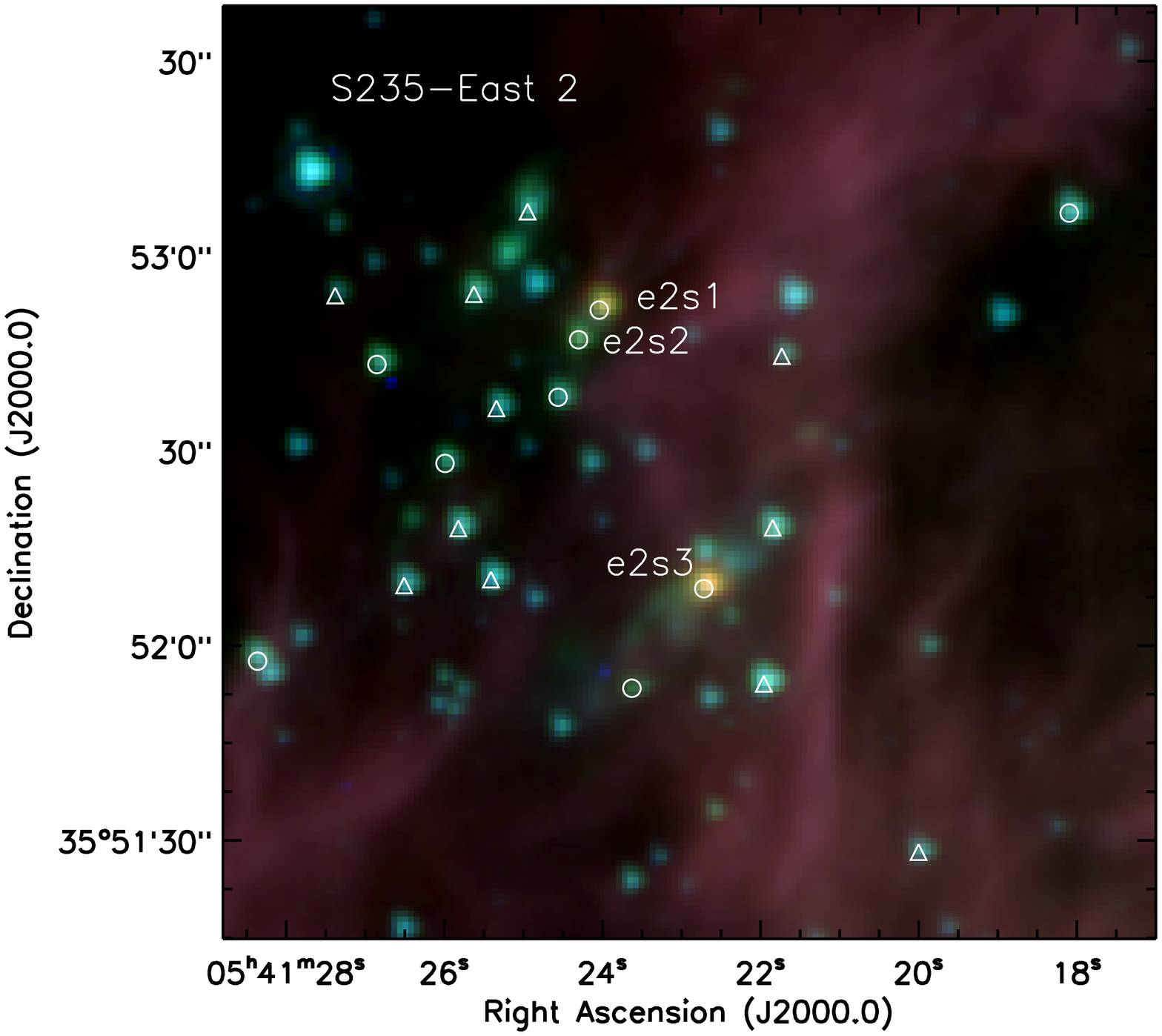}
\caption{IRAC 3-colour-composite (Ch4(red), Ch2(green) and Ch1(blue)) zoomed-in image of East~2 region (cf. Fig.~\ref{fig1}: dotted box).
Classified YSOs (Class 0/I in open circles and Class II in open triangles) are overlaid on the image.
The labeled sources are Class 0/I YSOs that are not detected in 2MASS JHK image. 
The extended ``green" (Ch2) region around the source labeled ``e2s3" may be due to shock-excited molecular 
H$_{2}$ emission and may be the signature of outflows associated with the source (see text).}
\label{fig5}
\end{figure*}

Figure~\ref{fig5} shows the East~2 region in IRAC 3-colour-composite image zoomed from Fig.~\ref{fig1} (dotted box).
We obtained 9 Class 0/I and 11 Class II sources in the East~2 region as displayed in the Fig.~\ref{fig5}.
Like in the case of East~1, in East~2 also we identified some interesting 
Class 0/I sources (labeled as `e2s1-e2s3', in Fig.~\ref{fig5}) that do not have 
JHK counterparts and are detected only in (all the four) IRAC bands. 
The labeled sources are listed in Table~\ref{tab3} with their positions and observed IRAC magnitudes.
The distance of inflection for this cluster is $d_{c}$ = 0.47 pc, nearly the same as that for 
the East 1 cluster.
The weighted mean values of the physical parameters obtained from SED modeling for all selected sources
are given in Table~\ref{tab4}. 
All these 3 selected sources are young and still passing through the accretion stage. 
Fig.~\ref{fig5} exhibits extended ``green (Ch2)" emission (about 0.27 pc in SE-NW direction) associated with the 
Class 0/I source ``e2s3". It could be possible that the source ``e2s3" is associated with outflows of shock-excited H$_{2}$. 
The SED modeling results (Table~\ref{tab4}) indicate that the source ``e2s3" is a massive but very young protostar not yet being able to 
excite an HII region. Our results suggest that ``e2s3" is a very interesting source in the S235 East~2 region which may be a new
candidate for a high mass protostellar object (HMPO) that is associated with outflows. 
The SED results are also discussed in the context of the IRAC ratio maps (see Sec. 3.2).
\subsubsection{S235A-B: combined IRAC and NIR photometry}
\begin{figure*}
\includegraphics[width=0.70\textwidth]{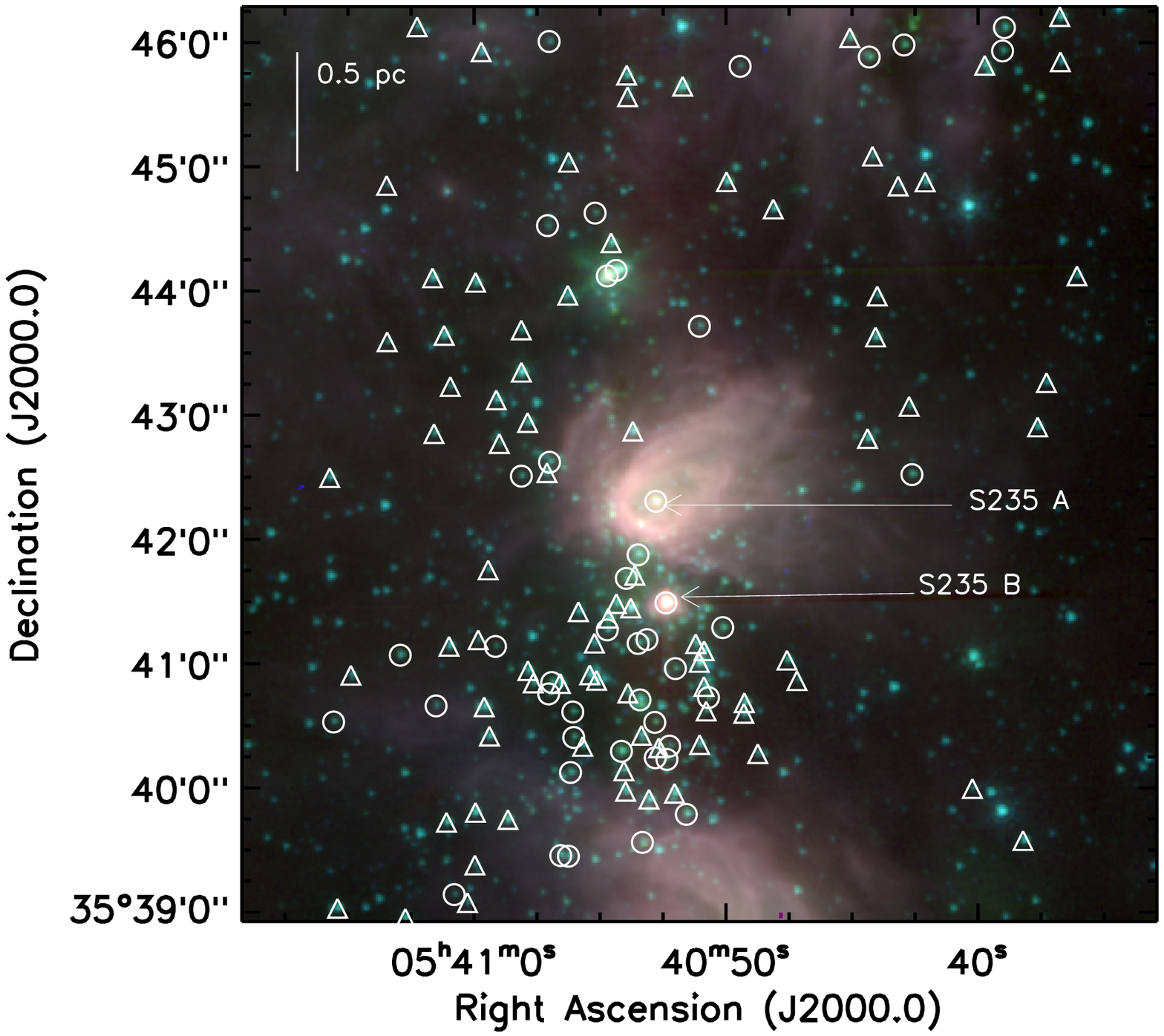}
\caption{IRAC 3-colour-composite (Ch4(red), Ch2(green) and Ch1(blue)) zoomed-in image of S235A-B region (shown in Fig.~\ref{fig1} by solid box).
$\alpha_{IRAC}$-classified YSOs (Class 0/I in open circles and Class II in open triangles) are overlaid on the image.
Image shows the horse-shoe shaped structure around the source S235A and a small nebulous region associated with S235B.
It appears that the structure is expanding towards south-east due to the expansion of HII region and may be triggering fresh star formation.}
\label{fig6}
\end{figure*}

\begin{figure*}
\includegraphics[width=0.60\textwidth]{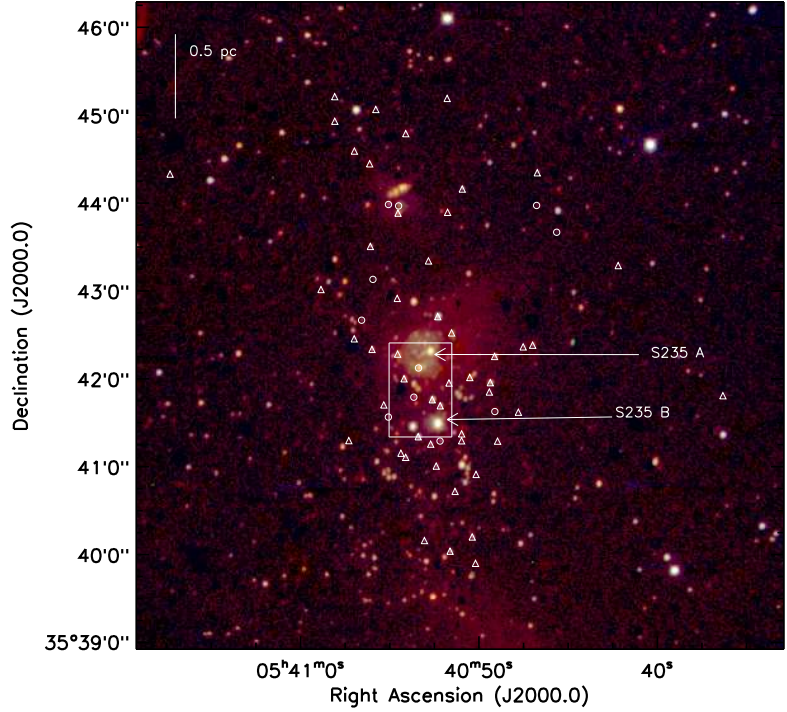}
\caption{The Mt. Abu NIR 3-colour-composite image (K(red), H(green) and J(blue)) of the vicinity of S235A-B region. 
Newly identified YSOs using HK$_{s}$+Ch1-2 criteria (see also Figure~\ref{fig8}) are overlaid on the image.
The Class I sources are shown by open circles and the Class II sources are in open triangles.}
\label{fig7}
\end{figure*}

\begin{figure*}
\includegraphics[width=0.565\textwidth]{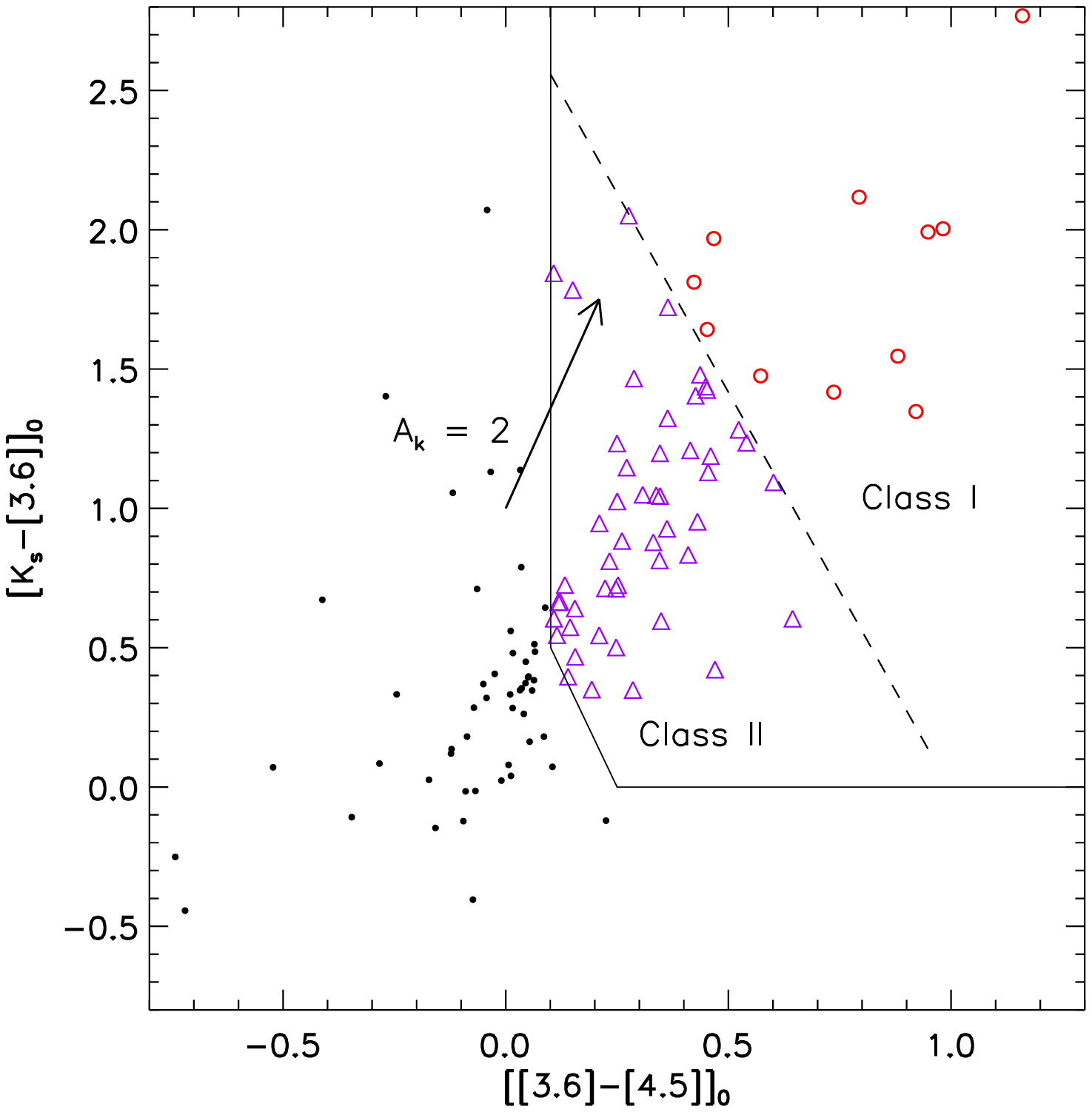}
\caption{De-reddened [K$_{s}$ - [3.6]]$_{0}$ $vs$ [[3.6] - [4.5]]$_{0}$ colour-colour diagram with the newly identified 
YSOs (Class I and Class II) in the S235A-B region. 
Open circles in red represent Class I and triangles in violet represent Class II sources and the rest are shown by 
filled circles. The region covered by the solid lines represents the location of YSOs. 
The dashed line marks the division between Class I and Class II sources. The extinction 
vector for A$_{K}$ = 2 mag is shown by the arrow, calculated using the average extinction law from \citet{Flaherty07}.} 
\label{fig8}
\end{figure*}

\begin{figure*}
\includegraphics[width=1.0\textwidth]{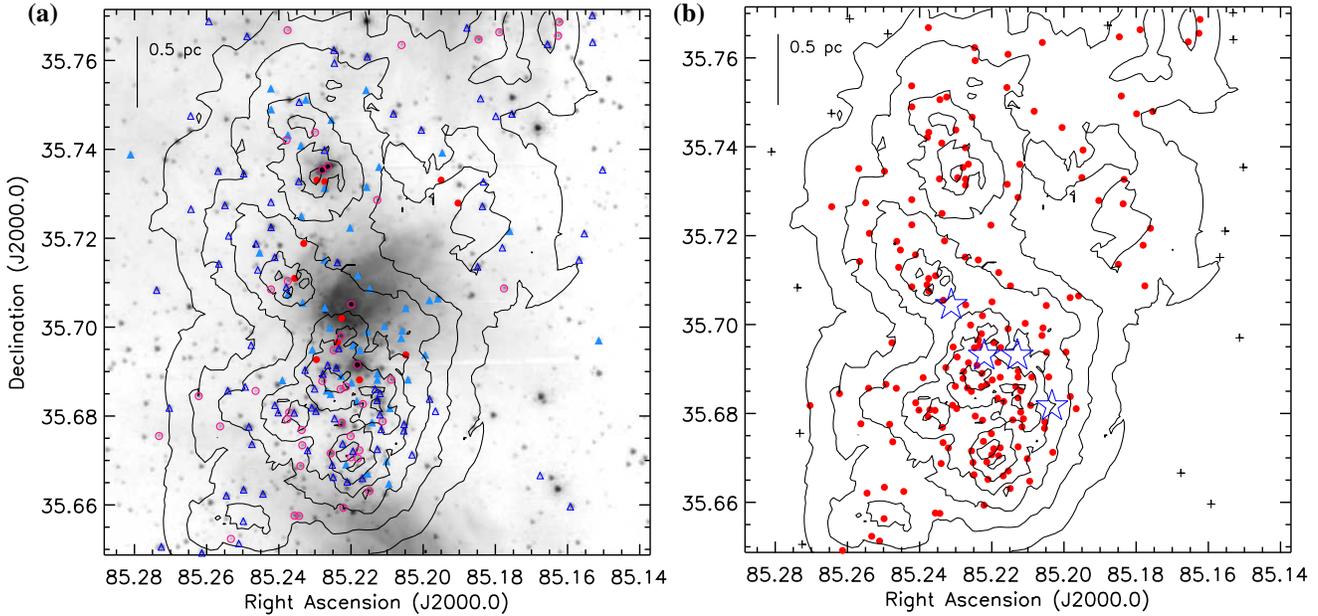}
\caption{a: Surface density map of YSOs overlaid on inverted IRAC Ch2 image. Circles and triangles represent Class I and Class II sources respectively.
Filled red circles show the Class I sources identified through H, K, CH1, CH2 bands and deep-pink open circles represent the Class 0/I sources selected from IRAC
four bands criteria. Similarly, Class II sources identified through H, K, CH1, CH2 bands are represented by filled light-blue
triangles and dark-blue open triangles
show the Class II sources identified through four IRAC bands criteria. The levels of surface density contours are 5,10, 20, 40, 70 and 120 YSOs/pc$^{2}$ from outer to
inner side. 
b: The distribution of all cluster members are shown by 
filled circles (red) for YSOs within $d_{c}$ $\leqslant$ 0.59 pc and plus symbols (black) for the YSOs outside $d_{c}$. 
The positions of dense clumps associated with the region (shown by blue star symbols) 
are detected from C$^{18}$O observations by \citet{Saito07}.} 
\label{fig9}
\end{figure*}

\begin{figure*}
\includegraphics[width=\textwidth]{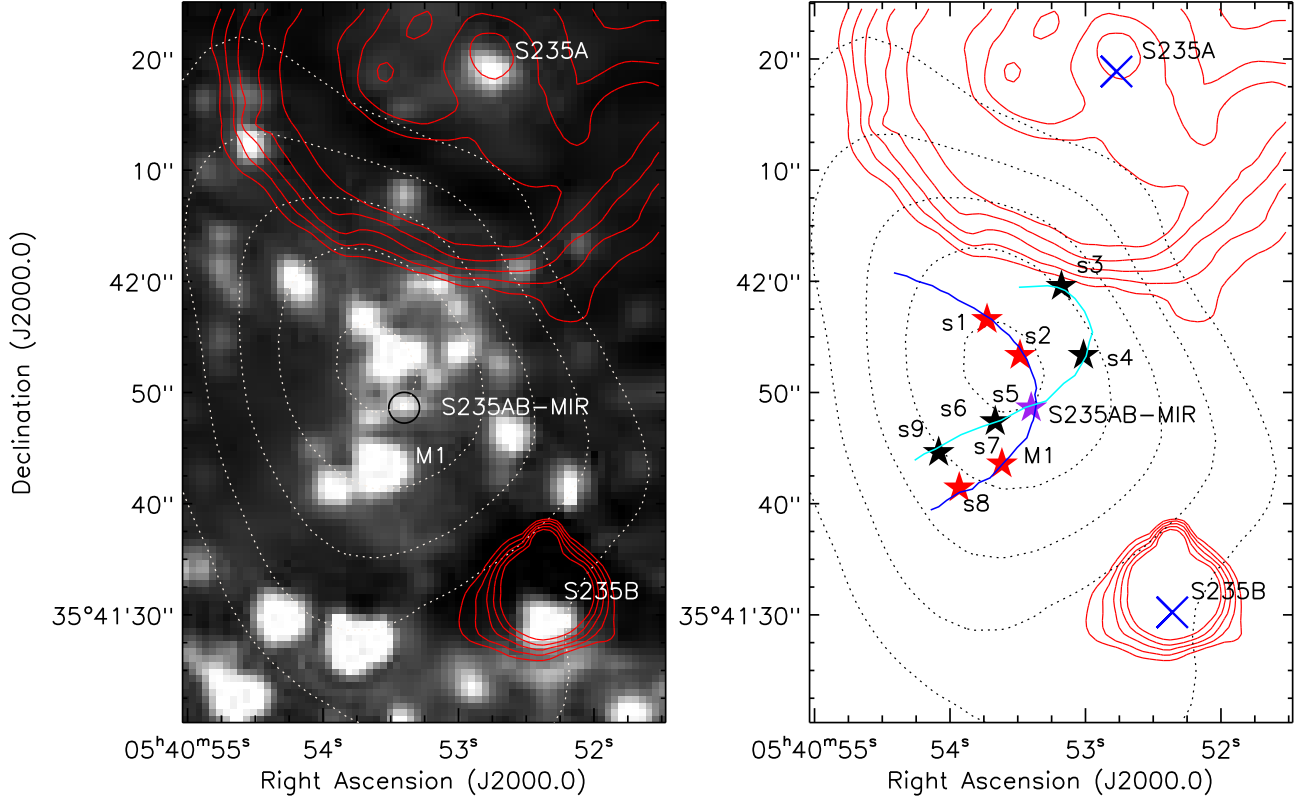}
\caption{The image on the left represents the zoomed-in IRAC ratio image of Ch2/Ch4 in the region marked by
the solid box in Fig.~\ref{fig7}. 
Right side image shows the positions of sources seen at least in any one of the IRAC bands with IRAC Ch4 flux contours (the full curves in red) 
overlaid (with flux levels between 420 and 1300 MJy/Sr).
Source S235AB-MIR is represented by a violet star symbol and the blue crosses show the position of sources S235A and S235B.
Public archival SCUBA 850 $\mu$m dust continuum contours (the dotted curves) of the region are overlaid on both the figures (with 
flux density levels between 1100 and 9900 mJy/beam). 
Two arc-like structures formed by sources s1 to s9 are shown in cyan (grey) and blue (black) curves in 
the right side figure (see text for details). 
One may notice that the sources S235AB-MIR (s5) and M1 (s7) are `members' of the arc formation. 
In order to show clearly the ring-structure we have used here a platescale of 0.61 arcsec/pixel.}
\label{fig10}
\end{figure*}

Fig.~\ref{fig6} represents the zoomed IRAC 3-colour-composite image in the the vicinity of the S235A-B region (solid box
in Fig.~\ref{fig1}). From the IRAC photometry (using the $\alpha_{IRAC}$ criteria), 
we obtained 128 YSOs (44 Class 0/I and 84 Class II) in this region  
whose locations are shown in the Figure. 
For IRAC sources that do not have good quality data in Ch3 and/or Ch4 bands, 
\citet{Gutermuth09} considered 2MASS data along with IRAC Ch1 and Ch2 bands and 
demonstrated the separation of protostars, stars with disks and disk-less photospheres
using the dereddened K$_{s}$ - [3.6] $vs$ [3.6] - [4.5] colour space obtained from the colour excess ratios and 
the reddening law of \citet{Flaherty07}. 
In Appendix B, we discuss the efficacy of the HK$_{s}$+Ch1-2 scheme in comparison with the $\alpha_{IRAC}$ criteria,   
for sources in S235AB region for which data on all IRAC bands and JHK are available.
Using Mt Abu NIR photometry, in combination with 
IRAC Ch1 and Ch2, we identified additional YSOs which are not detected in Ch3 and/or Ch4, following \citet{Gutermuth09}. 
While applying the HK$_{s}$+Ch1-2 criteria, we first converted the Mt Abu JHK(MKO-consortium) photometric magnitudes in to 
2MASS JHK$_{s}$ system following \citet{Legg06}. 
We used all the common sources in H, K, Ch1 and Ch2 bands (having positional coincidence within  
1 arcsec) to identify additional new YSOs in the region. 
We then eliminated the contaminants from the IRAC photometry, before applying the selection criteria of \citet{Gutermuth09}. 
The new YSOs numbering 61 (11 Class I and 50 Class II) are shown overlaid on 
the JHK-colour-composite image in Fig.~\ref{fig7} and thier positions and photometric magnitudes are 
listed in Table~\ref{tab5} (Class I) and Table~\ref{tab6} (Class II).  
Figure~\ref{fig8} shows the dereddened [K$_{s}$ - [3.6]]$_{0}$ and [[3.6] - [4.5]]$_{0}$ 
colour-colour diagram with demarcation lines between Class I and Class II sources.
Adding these sources to the IRAC identified sources, we get a total number of YSOs in the vicinity of the 
region S235A-B as 189, with 55 Class I and 134 class II sources. The Class III sources identified are 
4 from IRAC bands and 51 from HK$_{s}$+Ch1-2 bands making a total of 55. 
The IRAC analysis further gave 37 purely photospheric sources.\\\\
We have re-estimated the surface density for all the YSOs in the S235A-B region using a 5 arcsec grid size, as before.
Overlaid on the Ch2 band, the surface density contours are shown in Fig.~\ref{fig9}a, having a maximum surface density of
about 120 YSOs/pc$^{2}$. 
It is very clear that nearly all the Class I sources are lying within the surface density contours.  
We calculated the distance of inflection $d_{c}$ = 0.59 pc ($\sim$ 0.019$^{o}$) for the 
clustered YSOs shown clearly (with same symbols for both Class I and II) in Figure~\ref{fig9}b. 
We find 70\% of the YSOs to be present in clusters. When we used only the IRAC-identified YSOs in the region of S235A-B,
we get $d_{c}$ = 0.61 pc, giving  
about 86\% YSOs inside the clusters. This is in reasonable agreement with the NIR-IRAC combined result. 
The positions of the dense clumps (detected in C$^{18}$O measurements 
by \citet{Saito07}) associated with the S235A-B region are over-plotted in Fig.~\ref{fig9}b (as blue stars). The presence 
of C$^{18}$O emission supports the dense clustering of YSOs in the S235A-B region. 
We may identify 5-6 smaller clusters in S235A-B region (see Figure~\ref{fig9}b) 
with a minimum surface density of 40 YSOs/pc$^{2}$.\\\\
The S235A-B region was well studied by \citet{felli97,felli04,felli06}, who found evidences for 
fresh star formation in the region, through the interaction of an expanding classical HII region into an ambient molecular core.
Figure~\ref{fig10} (left-side) shows the zoomed-in image of the S235A \& B regions (solid
box in Fig.~\ref{fig7}).
Also shown in the figure on the right-side are some of the JHK/IRAC-identified sources (marked as red and black stars) labeled as 
s1-s9, over-plotted by the flux contours of Ch4 in full red curves and  
the SCUBA 850 $\mu$m dust continuum contours in dotted black curves (obtained from JCMT archival data base). 
The sky positions and the JHK/IRAC magnitudes of these labeled objects are listed in Table~\ref{tab7}. 
It is interesting to note that these 
sources seem to be most prominent in Ch2. We bring in to focus the one (or two) arc-like or semi-ring-like formation(s) 
made by the 
labeled sources (shown for better perspective by cyan (grey) and blue (black) curves in the right-side part of Figure~\ref{fig10}). 
Although one can 
notice a few more objects in the image (left-side in Figure~\ref{fig10}) which seem to be a part of the arc-like structure, 
these are not detected beyond uncertainties in the IRAC or JHK photometry.   
It may also be noticed that the `member-objects' of the arc-like structure are coincidental with 
the dense clumps identified by \citet{Saito07} as well as  
the 850 $\mu$m  dust continuum  in the S235A-B region (see right-side in Fig 10).\\\\
\citet{felli06} found a source S235AB-MIR (labeled as s5 in Fig 10) at the peak position of the 1.2 mm observations 
which has been detected in all the IRAC bands, except in Ch1.
It is interesting that the source S235AB-MIR (s5) is one of the `members' in the arc-like structure.
We did the SED modeling of the labeled sources using the model tool cited earlier. 
The source s5 has only 3 data points in IRAC (see Table~\ref{tab7}). It should be mentioned that modelling 
of SEDs with only 3 points may not be trustworthy due to the large degeneracy in 
the number of possible models. The degeneracy may be alleviated by 
constraining other input parameters such as visual extinction and distance. Therefore, s5 was modeled by adding 
the sub-mm-continuum data point from \citet{felli04}. 
The weighted mean values of the physical parameters obtained from the modeling are given in Table~\ref{tab8}.
The model-derived parameters show that the source S235AB-MIR is a young and accreting 
massive protostar that is not yet 
able to excite an HII region. Also, the results of SED modeling of the source M1 (labeled as s7 in Fig 10) show that it is a low-mass star 
but not highly evolved (see Table~\ref{tab8}).
\citet{felli06} reported that the radio source VLA-1 coincides with the source M1 and that it could be a B2-B3 star with a UCHII region.
But, our SED modeling results of M1 (s7) are not consistent with the suggestion of \citet{felli06}.\\\\
The right-side of Fig.~\ref{fig10} shows the horse-shoe shaped structure in the contours (full in red) of IRAC Ch4 band  
towards the south, forming the outer boundary of the expanding HII region.
The semi-ring-like structure of sources is just located at the interface between the horse-shoe shaped structure 
and the peak of the dust continuum. It may hence be conjectured that 
triggered star formation by the ``collect and collapse'' process \citep{Elmegreen77} could be responsible for the 
existence of these IRAC sources.
There is also the possibility for the formation of stars due to the propagation of a shock wave from the HII region 
which may lead to the contraction of pre-existing dense molecular clumps. This possibility is supported by the detection
of dense C$^{18}$O emission clumps in the S235A and S235B regions \citep{Saito07} (see also \citep{felli06}).
\citet{li01} and \citet{li02} made a simulation for the formation of arc/ring-like structures and multiple-star 
formation in magnetically sub-critical clouds. 
They demonstrated that a magnetically sub-critical cloud breaks up into fragments as multiple magnetically 
supercritical clumps leading to the formation of a new generation of stars in 
small clusters (in a arc/ring-like formation). 
A supportive observational evidence was earlier reported by \citet{kumar03} in the source IRAS 22134+5834.
Interestingly, the distances between the immediate neighbouring ``members" in the present case 
($\sim$ 0.04 pc) are very close to the theoretically predicted values for the cores (viz. 0.05 pc). 
Two difficulties may however be noted: firstly, the theory deals with relatively quiescent situations; 
secondly, the protostar ``members" should be coeval. As may be seen from Fig.~\ref{fig10}, 
the location of the ring-like structure is 
farther from the HII region shock front than those situated closely along the front (see also Fig 2 of \citet{felli06})
and hence may be relatively quiescent.
The age spreads for clusters are typically a few Myr \citep{schulz05}. In the present case, the
model-derived ages for the ``members" show a range of values between 10$^{4}$ to
10$^{6}$ years. Considering the uncertainties in the model-derived ages and the lack of magnetic field measurements,   
our suggestion of magnetic field mediated cluster formation remains a speculation.

In the next section we examine the three cluster regions in the light of the ratio images of the IRAC bands.   

\subsection{IRAC Ratio Maps}
\begin{figure*}
\includegraphics[width=0.65\textwidth]{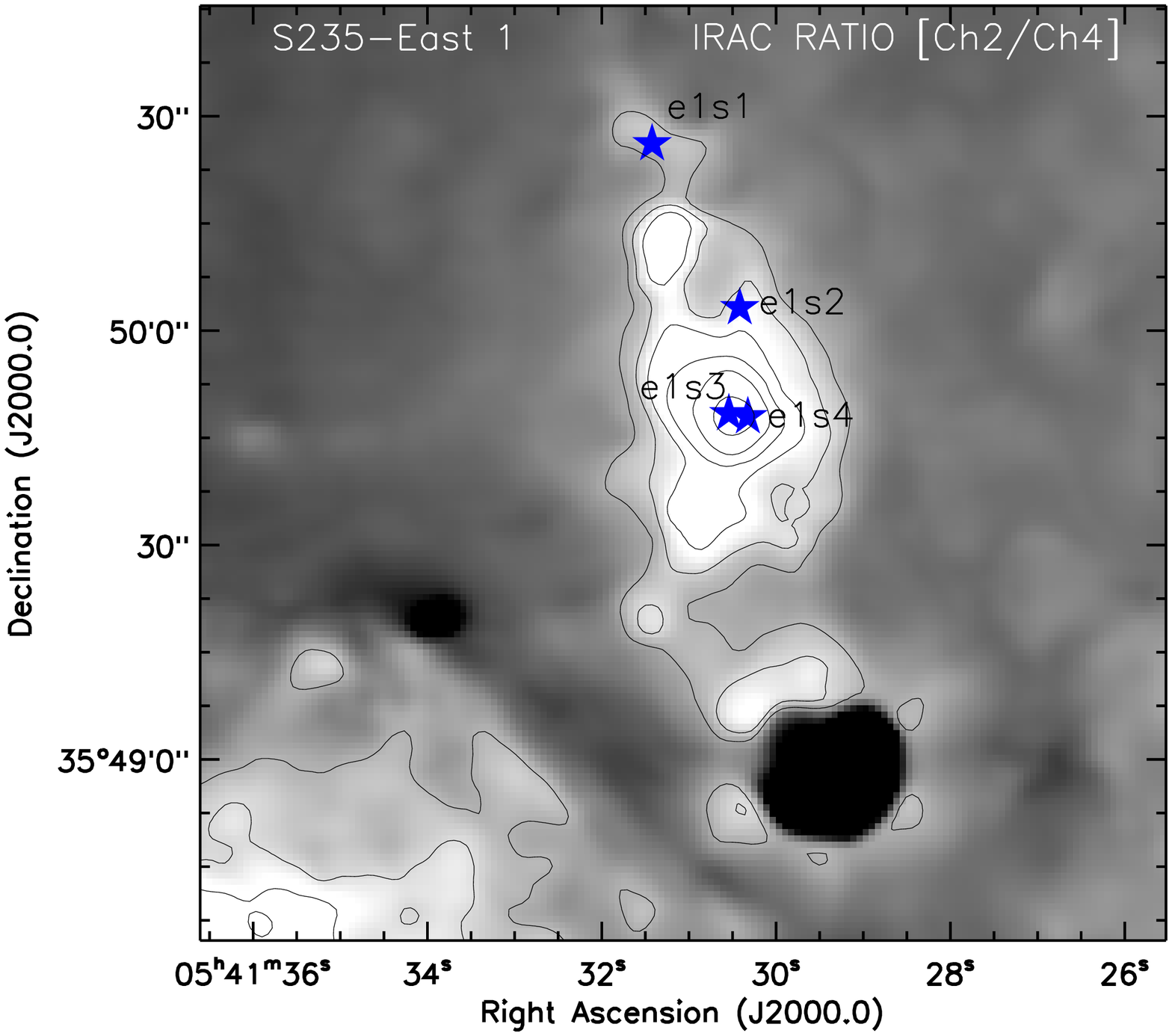}
\caption{IRAC ratio map of Ch2/Ch4 in East~1 region revealing bright regions. Contours of the ratio map of Ch2/Ch4 are over-plotted 
with labeled sources (blue stars)  from Fig.~\ref{fig4}.
The contour levels are min 0.04 and max 0.68.}
\label{fig11}
\end{figure*}

\begin{figure*}
\includegraphics[width=0.65\textwidth]{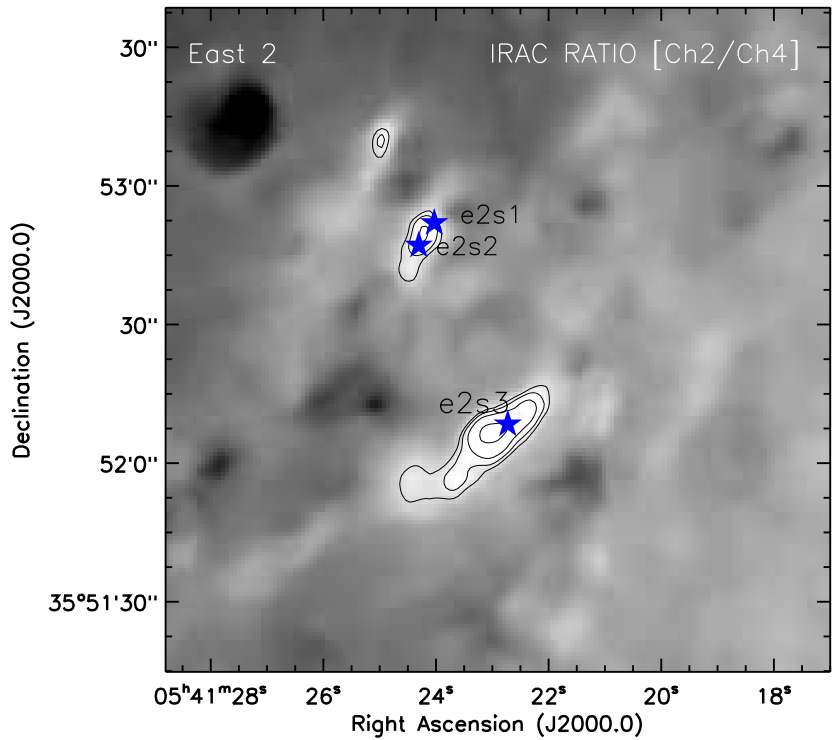}
\caption{IRAC ratio map of Ch2/Ch4 in East~2 region revealing bright regions.
Contours of the ratio map of Ch2/Ch4 are over-plotted with labeled sources (blue stars) from
Fig.~\ref{fig5}.
The contour levels are min = 0.04 and max = 0.08. Source ``e2s3" may be associated with an outflow due to shock-excited H$_{2}$ emission. 
The extension of bright region (end-to-end) associated with source ``e2s3" is about 0.27 pc 
in SE-NW direction.}
\label{fig12}
\end{figure*}

\begin{figure*}
\includegraphics[width=0.65\textwidth]{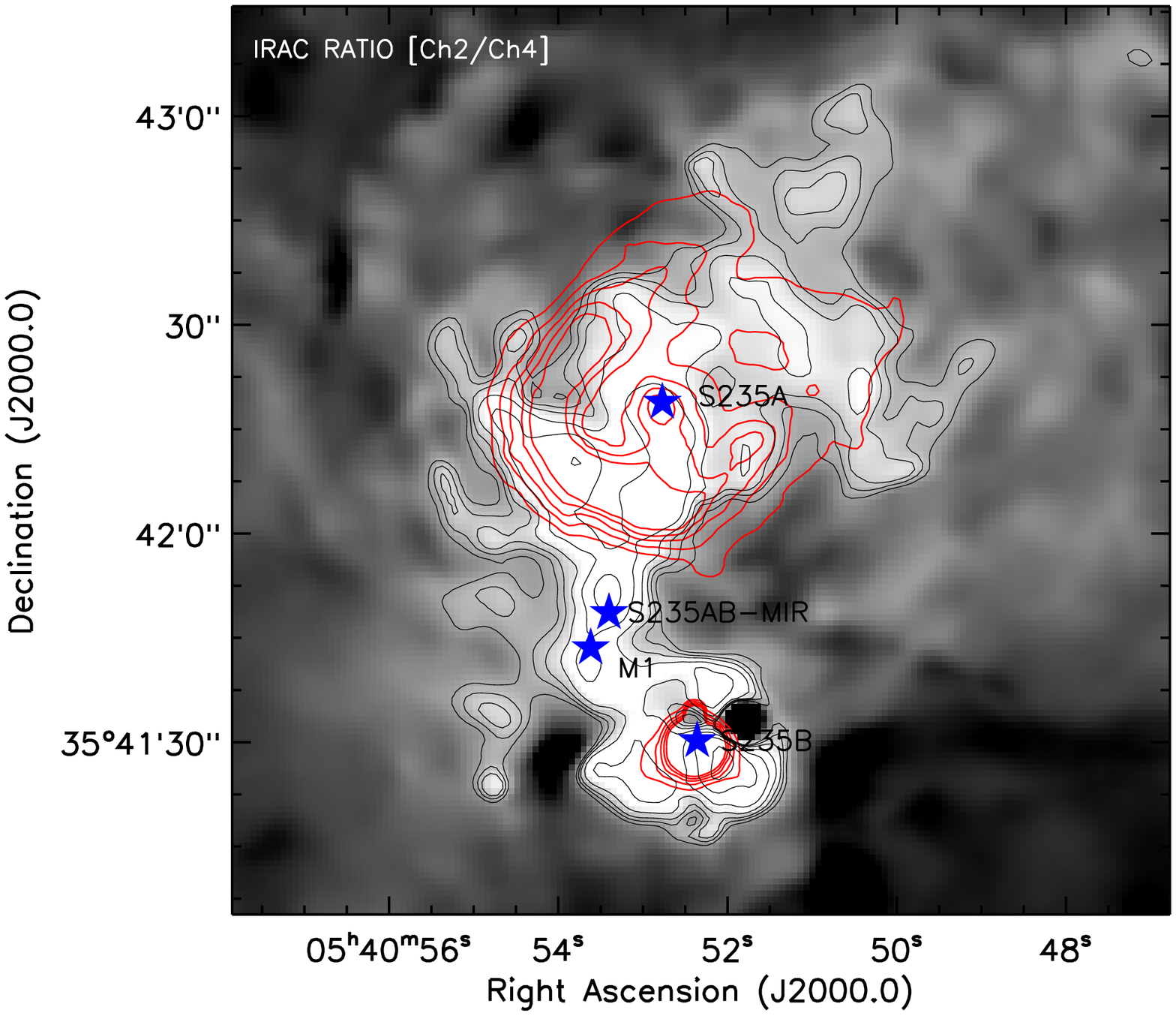}
\caption{IRAC ratio map of Ch2/Ch4 in the S235A-B region revealing bright regions.
Contours of IRAC Ch4 are over-plotted in red colour, with same levels as shown in Fig.~\ref{fig10}.
Bright emission regions within red contours are probably due to Br$\alpha$ emission. The bright regions outside towards south
of the red contours are probably due to molecular H$_{2}$ emission.
Also, the contours of the IRAC ratio map of Ch2/Ch4 are overlaid with levels of min=0.04 and max=0.77. The position of important sources are
marked by blue star symbols on the image. It is noticed that the S235B is also associated with a 
bright regions that could be due to Br$\alpha$ emission due to the presence of an HII region.}
\label{fig13}
\end{figure*}

Several authors utilised the ratios of 
IRAC bands to identify some atomic and molecular lines/bands/features occurring within the bands 
(e.g., \citet{Smith05,Povich07,Neufeld08,dewangan10,lokesh10}). 
In particular, since the IRAC Ch2 (4.5 $\mu$m) band does not include any PAH features, it may be used as a 
reference to trace out the PAH features when ratio-ed with the other 3 IRAC bands. Further, Ch2 
includes in it the emission from HI Br$\alpha$ (4.05 $\mu$m) and hence can well trace out HII regions, compared to other HI lines 
in the other IRAC bands (see \citet{Smith05}).  
We used the IRAC ratio Ch2/Ch4 to find out the emission regions in the S235 complex.
In the ratio image of Ch2/Ch4, the brighter regions 
indicate emission regions from higher excitations from H$_{2}$ and the darker regions indicate PAH emission. 
This trend is reversed in the image of Ch4/Ch2 (i.e., bright regions show PAH emission and 
dark regions show the H$_{2}$ emission). It may be noted here that the ratio maps are only 
indicative; until/unless confirmed by independent spectroscopic evidence.\\\\ 
In order to make the ratio maps, point sources from all the IRAC images are removed by using an extended 
aperture of 12.2 arcsec and a sky-annulus of 12.2-24.4 arcsec in IRAF/DAOPHOT software \citep{Reach05}.
Then, these residual frames are subjected to median-filtering with a width of 9 pixels and smoothing by 3$\times$3 pixels 
using the BOXCAR algorithm in IRAF (as in \citet{Povich07}). 
IRAC zoomed-in ratio map (Ch2/Ch4) of the East~1 region is shown in Fig.~\ref{fig11}.
The smallest value of the ratio Ch2/Ch4 that may be trusted is 0.04.
We found a very strong bright region corresponding to East~1 in the ratio map which could be due to 
H$_{2}$ emission.
The results of SED modeling of important sources (i.e., e1s1, e1s2, e1s3 and e1s4) in the region show 
the lack of an associated HII region. But these sources are young, still accreting material and may be 
associated with outflows, especially the Class 0/I sources e1s1, e1s3 and e1s4.
The IRAC zoomed-in ratio map (Ch2/Ch4) of the East~2 region is shown in Fig.~\ref{fig12}.
The ratio map shows an extended bright emission possibly 
associated with the source ``e2s3".
This could represent the outflow lobes of a possible jet from the source due to shock-excited molecular H$_{2}$ emission.
The size of the bright region (or ``jet") is about 0.27 pc in SE-NW direction (in the contour level of 0.046).
The SED modeling of the source ``e2s3" shows that it is a young massive protostar still accreting the material (see
Table~\ref{tab4}).
Our results are only indicative/speculative about the nature of these sources; therefore, it requires further study at 
longer wavelengths i.e., sub-mm/mm.\\\\
Fig.~\ref{fig13} shows the IRAC ratio map of Ch2/Ch4 in the S235A-B region. We find a very strong 
bright emission in the Ch2/Ch4 map in the S235A-B region.
The IRAC horse-shoe shaped structure seen in Ch4 band is over-plotted as contours in Fig.~\ref{fig13}.
We may infer from the ratio map that the Br$\alpha$ emission is present within the horse-shoe structure around S235A and represents
an HII region. But bright emission outside of the IRAC horse-shoe structure may be due to the H$_{2}$ emission.
This interpretation is consistent with the narrow band Br$\gamma$ and H$_{2}$ imaging of the region by \citet{felli97}.
We also found that the source S235B is associated with Br$\alpha$ emission due to the presence of an HII region.
Recently, using spectroscopy, \citet{boley09} reported that S235B is a source that belongs to a rare class of 
early-type Herbig Be stars. SED modeling of the source S235B (see Table~\ref{tab8}) shows that it is an intermediate mass young star and our 
model results are consistent with the \citet{boley09} results.
\section{Conclusions}
\label{sec:conc}
The important conclusions of this work are as follows:
\begin{enumerate}

\item{{\it Spitzer}-IRAC photometry of the S235 complex revealed 86 Class 0/I and 144 Class II YSOs. 
About 73\% of these are present in clusters concentrated in the regions, East~1, East~2, Central and 
the vicinity of S235A-B, with surface densities of 20-30 YSOs/pc$^{2}$;}

\item{A total of 189 YSOs (55 Class I and 134 Class II) are identified in the vicinity of the S235A-B 
region using a combination of NIR and IRAC bands. About 70\% of these sources are present in 5-6 clusters 
with maximum YSO surface density of about 120 YSOs/pc$^{2}$;}

\item{YSOs in a arc-like formation are found in the interface between the horse-shoe 
structure and sub-mm dust continuum emission in the S235A-B region. Also, one embedded 
source S235AB-MIR is detected which forms a part of the arc-like structure. 
SED modeling shows that this source is a young, massive star that is still accreting 
material. The arc-like structure may be speculated as an evidence for magnetically super-critical 
collapse. The SED modeling of one more member of the arc structure, M1, shows that it 
is a low-mass star, relatively young in its evolution;}

\item{The HII region associated with S235A is traced by the presence of Br$\alpha$ emission within 
the IRAC observed horse-shoe structure in the IRAC ratio Ch2/Ch4 map. Outside of the horse-shoe 
structure, the map indicates molecular H$_{2}$ emission;}

\item{Br$\alpha$ emission is also found around S235B using the ratio map Ch2/Ch4. The map reveals the 
presence of an ionised region around it and the SED modeling shows that the source S235B is a young 
intermediate mass star;}

\item{New Class 0/I sources are identified in the East~1 and East~2 regions, which do not have NIR JHK 
counterparts. IRAC colour-composite image reveals the presence of shock-excited H$_{2}$ emission in 
both the regions. This is confirmed in IRAC ratio map of Ch2/Ch4, which reveals that the source ``e2s3" in 
the East~2 region may be associated with shock-excited H$_2$ emission outflow or jet. The SED modeling of the 
source ``e2s3" in the East~2 region indicates that it is a massive but very young protostar (a candidate HMPO) not 
yet being able to drive an HII region.}
\end{enumerate}

\section*{Acknowledgments}

The research work is supported by the Department of Space, Government of India at PRL. 
We thank V. Venkataraman and the Mt Abu Observatory staff for observations using NICS.
A major part of this work is based on archival data from observations made with the 
Spitzer Space Telescope, which is operated by the Jet Propulsion Laboratory, 
California Institute of Technology under a contract with NASA. The authors express their 
gratitude to the anonymous referee for helpful comments.

\newpage

\appendix
\section{`Red-sources' among scattered YSOs}
We applied the `red-source' criteria of \citet{Robitaille08} to our identified scattered sources outside the clusters and 
have estimated nearly 7\% of the sources as the possible AGB contaminants (cf. \citet{lokesh10}). 
While the best way to distinguish between the YSOs and AGB stars is by spectroscopy, \citet{Robitaille08} show that 
the two classes may be well separated in the [8-24] colour in which the YSOs appear redder (see also \citet{Whitney08}). 
In our case, due to the absence of the 24 $\mu$m band observations, we may have estimated the `red-source' contamination 
due only to the `extreme AGBs' (having very red near-IR colours, 
as pointed out by \citet{Robitaille07,Robitaille08} and the references therein).
\section{Comparison between the classification criteria}
For sources in S235AB region which have both IRAC and JHK data, a comparison of classification was done using 
the criteria based on $\alpha_{IRAC}$ and HK$_{s}$+Ch1-2 magnitudes.   
Of the 28 sources identified as Class I using the HK$_{s}$+Ch1-2 criteria only 8 are identified as Class 0/I 
using the $\alpha_{IRAC}$ criteria (and 19 are identified as Class II). In comparison, of the 24 sources identified as Class II
using the HK$_{s}$+Ch1-2 criteria, 20 are identified as Class II in the $\alpha_{IRAC}$ criteria. But when we consider 
all the YSOs (Class I and II), the two methods match reasonably well (to about 90\%).  
We made a similar exercise on the data in Table 1 of \citet{hart05} for Taurus pre-main-sequence sources and found the same  
trends as in the present case. This may indicate that HK$_{s}$+Ch1-2 criteria are capable of classifying YSOs (both Class I and II 
together) but tend to over-estimate the Class I sources in comparison to the IRAC-based criteria.   
\section{Tables of identified sources}
\begin{table*}
\centering
\scriptsize
\caption{{\it Spitzer}-IRAC 4-channel photometry (in mag) of the Class 0/I YSOs identified in 
S235 complex (see text for details). NIR magnitudes are also given -
entries in italics are from Mt Abu observations (MKO system) and the rest are from 2MASS (JHK$_{s}$).}
\label{tab1}
\begin{tabular}{lcccccccccccccccccccr}
\hline
Object      &	RA [2000] &  Dec [2000] &  J    & H &  K   &  Ch1    &   Ch2  &  Ch3   &  Ch4 &$\alpha_{IRAC}$\\
\hline       		   	                                      	      	      	 	 	 	   	   	     	    	     	      		 
      1        & 5:40:33.05& 35:47:09.9&     17.14$\pm$0.19&	    -		   &	  15.36$\pm$0.17  &  14.22$\pm$0.01 &	13.79$\pm$0.01  &   13.32$\pm$0.11  & 11.83$\pm$0.08 &  -0.13$\pm$0.49     \\
      2        & 5:40:38.95& 35:46:07.2&     -  	   &	    -		   &	  -		  &  12.74$\pm$0.01 &	11.93$\pm$0.01  &   11.07$\pm$0.03  & 10.38$\pm$0.06 &  -0.11$\pm$0.23     \\
      3        & 5:40:39.03& 35:45:55.9&      - 	   &	    -		   &	  -		  &  13.48$\pm$0.02 &	12.97$\pm$0.02  &   12.12$\pm$0.04  & 10.91$\pm$0.04 &   0.17$\pm$0.24     \\
      4        & 5:40:40.74& 35:57:34.1&     16.35$\pm$0.10&	    15.02$\pm$0.07 &	  14.52$\pm$0.06  &  13.69$\pm$0.01 &	13.37$\pm$0.01  &   12.71$\pm$0.04  & 11.45$\pm$0.06 &  -0.22$\pm$0.39     \\
 \it  5        & 5:40:42.62& 35:42:31.3& \it 16.93$\pm$0.10&	\it 15.40$\pm$0.04 &  \it 14.34$\pm$0.05  &  13.26$\pm$0.01 &	12.61$\pm$0.01  &   12.08$\pm$0.03  & 10.79$\pm$0.04 &  -0.06$\pm$0.27     \\
      6        & 5:40:42.94& 35:45:58.8&     -  	   &	    -		   &	  -		  &  13.72$\pm$0.02 &	12.52$\pm$0.01  &   11.88$\pm$0.08  & 11.09$\pm$0.17 &   0.08$\pm$0.39     \\
      7        & 5:40:44.14& 35:54:59.2&     -  	   &	    14.83$\pm$0.17 &	  12.81$\pm$0.05  &  10.02$\pm$0.01 &	 8.87$\pm$0.00  &    7.88$\pm$0.01  &  6.91$\pm$0.02 &   0.72$\pm$0.28     \\
\hline          												    														      
\end{tabular}													    														      
\end{table*}													    														      
														    														      
\begin{table*}													    														      
\centering													    														      
\scriptsize																											      
\contcaption{}													    														      
\begin{tabular}{lcccccccccccccccccccr}																								      
\hline																												      
Object      &	RA [2000] &  Dec [2000] &  J & H &  K  &  Ch1    &   Ch2  &  Ch3   &  Ch4 &$\alpha_{IRAC}$\\															      
\hline       		   	                                      	      	   																		      
  81   &     5:41:30.55&      35:49:48.5&     -             &	     -              &	   -               &	    10.24$\pm$0.01 &	    8.88$\pm$0.00  &	   7.74$\pm$0.00 &	 7.01$\pm$0.00 &      0.87$\pm$0.54   \\	
  82   &     5:41:31.41&      35:50:26.1&     -             &	     -              &	   -               &	    13.29$\pm$0.02 &	    11.3$\pm$0.01  &	  10.06$\pm$0.01 &	 9.10$\pm$0.02 &      1.88$\pm$0.81    \\
  83   &     5:41:32.58&      35:49:19.4&     -             &	     -              &	   15.14$\pm$0.12  &	    12.95$\pm$0.01 &	    12.2$\pm$0.01  &	  11.43$\pm$0.05 &	10.34$\pm$0.07 &      0.16$\pm$0.04   \\
  84   &     5:41:36.22&      35:49:44.1&     -             &	     -              &	   -               &	    14.76$\pm$0.03 &	   13.52$\pm$0.01  &	  12.29$\pm$0.07 &	11.05$\pm$0.11 &      1.43$\pm$0.23   \\
  85   &     5:41:37.69&      35:45:34.3&     17.02$\pm$0.18&	     15.47$\pm$0.11 &	   14.11$\pm$0.06  &	    12.95$\pm$0.01 &	   11.78$\pm$0.01  &	  10.94$\pm$0.01 &	 9.85$\pm$0.02 &      0.66$\pm$0.24   \\
  86   &     5:41:38.06&      35:58:09.3&     -             &	     -              &	   -               &	    15.14$\pm$0.03 &	    14.7$\pm$0.03  &	  14.09$\pm$0.16 &	 12.7$\pm$0.18 &     -0.03$\pm$0.40   \\
\hline  																										      
\end{tabular}																										      
\end{table*}

\begin{table*}
\centering
\scriptsize
\caption{{\it Spitzer}-IRAC 4-channel photometry (in mag) of the Class II YSOs identified in S235 
complex (see text for details). NIR magnitudes are also given -
entries in italics are from Mt Abu observations (MKO system) and the rest are from 2MASS (JHK$_{s}$).}
\label{tab2}
\begin{tabular}{lcccccccccccccccccccr}
\hline
Object      &	RA [2000] &  Dec [2000] &  J & H &  K   &  Ch1    &   Ch2  &  Ch3   &  Ch4 &$\alpha_{IRAC}$\\
\hline       		   	                                      	      	      	 	 	 	   	   	     	    	      		 
       1    &  5:40:20.14  &   35:55:23.0 &	 14.57$\pm$0.03  &	13.37$\pm$0.04  &      12.78$\pm$0.03	&   12.12$\pm$0.00   &    11.84$\pm$0.00    &	11.33$\pm$0.02  &     10.70$\pm$0.03   &    -1.17$\pm$0.13     \\ 
       2    &  5:40:30.50  &   35:45:44.7 &	 16.51$\pm$0.11  &	15.53$\pm$0.10  &      15.31$\pm$0.15	&   14.55$\pm$0.01   &    14.25$\pm$0.02    &	13.88$\pm$0.09  &     13.09$\pm$0.12   &    -1.17$\pm$0.20     \\ 
       3    &  5:40:34.86  &   35:55:05.5 &	 16.93$\pm$0.16  &	14.90$\pm$0.06  &      13.57$\pm$0.04	&   11.35$\pm$0.00   &    10.52$\pm$0.00    &	 9.97$\pm$0.01  &      9.33$\pm$0.02   &    -0.57$\pm$0.22     \\ 
       4    &  5:40:35.28  &   35:46:21.8 &	 16.41$\pm$0.10  &	14.97$\pm$0.06  &      14.61$\pm$0.08	&   13.73$\pm$0.01   &    13.35$\pm$0.01    &	12.66$\pm$0.04  &     12.07$\pm$0.06   &    -0.87$\pm$0.14     \\ 
       5    &  5:40:35.94  &   35:55:29.2 &	 14.77$\pm$0.04  &	13.23$\pm$0.03  &      12.30$\pm$0.02	&   10.97$\pm$0.00   &    10.54$\pm$0.00    &	10.16$\pm$0.01  &      9.46$\pm$0.01   &    -1.12$\pm$0.09     \\ 
       6    &  5:40:36.07  &   35:44:07.3 &		 -	 &		-	&	       -	&   13.17$\pm$0.01   &    12.96$\pm$0.01    &	12.66$\pm$0.03  &     12.08$\pm$0.06   &    -1.58$\pm$0.15     \\ 
       7    &  5:40:36.73  &   35:45:50.7 &		 -	 &		-	&	       -	&   13.14$\pm$0.01   &    12.87$\pm$0.01    &	12.37$\pm$0.04  &     12.08$\pm$0.06   &    -1.57$\pm$0.15     \\
\hline          												    
\end{tabular}													    
\end{table*}													    
														    
\begin{table*}													    
\centering													    
\scriptsize
\contcaption{}													    
\begin{tabular}{lcccccccccccccccccccr}
\hline
Object      &	RA [2000] &  Dec [2000] &  J & H &  K  &  Ch1    &   Ch2  &  Ch3   &  Ch4 &$\alpha_{IRAC}$\\
\hline  																			       
  	 64    &  5:40:55.69  &   35:40:20.1 &  	   -	    &		  -	   &		 -	   &   13.38$\pm$0.01	&    12.79$\pm$0.01    &   12.28$\pm$0.04  &	 11.73$\pm$0.11   &    -0.96$\pm$0.11	  \\ 
    \it  65    &  5:40:55.86  &   35:41:25.1 &  \it 16.24$\pm$0.00  &  \it 14.80$\pm$0.04  &  \it 13.88$\pm$0.04   &   12.51$\pm$0.01	&    12.05$\pm$0.01    &   11.81$\pm$0.09  &	 11.27$\pm$0.08   &    -1.47$\pm$0.09	  \\ 
    \it  66    &  5:40:56.25  &   35:45:02.2 &  	   -	    &  \it 15.91$\pm$0.04  &  \it 14.87$\pm$0.06   &   13.13$\pm$0.01	&    12.43$\pm$0.01    &   12.05$\pm$0.07  &	 11.04$\pm$0.15   &    -0.52$\pm$0.20	  \\ 
    \it  67    &  5:40:56.28  &   35:43:58.0 &  \it 16.24$\pm$0.00  &  \it 14.22$\pm$0.04  &  \it 13.06$\pm$0.03   &   11.93$\pm$0.01	&    11.43$\pm$0.01    &   11.10$\pm$0.04  &	 10.68$\pm$0.08   &    -1.43$\pm$0.12	  \\ 
  	 68    &  5:40:56.58  &   35:40:50.5 &  	   -	    &		  -	   &		 -	   &   12.12$\pm$0.00	&    11.64$\pm$0.00    &   10.96$\pm$0.01  &	 10.51$\pm$0.02   &    -0.93$\pm$0.18	  \\ 
    \it  69    &  5:40:57.11  &   35:42:32.3 &  \it 15.25$\pm$0.00  &  \it 14.14$\pm$0.03  &  \it 13.35$\pm$0.04   &   11.97$\pm$0.01	&    11.47$\pm$0.01    &   11.33$\pm$0.12  &	  9.66$\pm$0.10   &    -0.31$\pm$0.67	  \\ 
    \it  70    &  5:40:57.65  &   35:40:50.7 &  \it 16.06$\pm$0.00  &  \it 14.97$\pm$0.03  &  \it 14.31$\pm$0.05   &   13.04$\pm$0.01	&    12.55$\pm$0.01    &   12.04$\pm$0.04  &	 11.53$\pm$0.08   &    -1.08$\pm$0.08	  \\
\hline 
\end{tabular}
\end{table*}

\begin{table*}
\centering
\caption{Selected sources in S235 East~1 and East~2 regions; the numbers in parentheses in the Object column are 
common sources from Table~\ref{tab1} designations (see Figs.~\ref{fig4} \& ~\ref{fig5} and text). 
All sources are Class 0/I and do not have JHK counterparts.}
\label{tab3}
\begin{tabular}{lcccccccccccccccr}
\hline
  Object   &   RA [2000]&  Dec [2000]&  Ch1&   Ch2&  Ch3&  Ch4&$\alpha_{IRAC}$	\\
\hline          							             
  e1s1(82) & 5:41:31.41 & 35:50:26.1 &  13.29$\pm$0.02 &  11.30$\pm$0.01 & 10.06$\pm$0.01 &  9.10$\pm$0.02 &  1.88$\pm$0.81   \\ 
  e1s2(80) & 5:41:30.41 & 35:50:03.4 &  12.17$\pm$0.01 &  10.54$\pm$0.01 &  9.56$\pm$0.01 &  8.73$\pm$0.03 &  1.03$\pm$0.65   \\
  e1s3(81) & 5:41:30.55 & 35:49:48.5 &  10.24$\pm$0.01 &   8.88$\pm$0.00 &  7.74$\pm$0.00 &  7.01$\pm$0.00 &  0.87$\pm$0.54   \\
  e1s4(79) & 5:41:30.32 & 35:49:48.1 &  10.01$\pm$0.01 &   8.75$\pm$0.00 &  7.72$\pm$0.00 &  6.78$\pm$0.00 &  0.85$\pm$0.36   \\
  e2s1(70) & 5:41:24.04 & 35:52:51.7 &  13.06$\pm$0.03 &  11.12$\pm$0.01 &  9.40$\pm$0.01 &  7.94$\pm$0.02 &  3.04$\pm$0.59   \\
  e2s2(71) & 5:41:24.30 & 35:52:47.1 &  13.27$\pm$0.03 &  11.71$\pm$0.01 & 10.48$\pm$0.04 &  9.26$\pm$0.07 &  1.72$\pm$0.43   \\
  e2s3(67) & 5:41:22.72 & 35:52:08.8 &  12.33$\pm$0.03 &  10.33$\pm$0.01 &  8.54$\pm$0.01 &  6.97$\pm$0.01 &  3.31$\pm$0.57   \\
\hline 
\end{tabular}
\end{table*}

\begin{table*}
\scriptsize
\centering
\caption{Weighted mean values of the physical parameters 
derived from SED modeling of the labeled sources from Table~\ref{tab3} in S235 East~1 and East~2 regions (see text).
$\chi^{2}$ values are given per data point.}
\label{tab4}
\begin{tabular}{lcccccccccr}
    \hline 
Source  &$\chi^{2}$       & Age 	          & $M_*$	  &	$T_*$	    & $L_*$	      &  M$_{env}$	  &  $\dot{M}_{env}$		&  M$_{disk}$	    & $\dot{M}_{disk}$  	 &  No. of models \\
Name    &                & log(yr)	  & M$_{\odot}$   &log(T(K))	    & log(L$_{\odot}$)& log(M$_{\odot}$ ) & log(M$_{\odot}$ yr$^{-1}$)  & log(M$_{\odot}$ ) & log(M$_{\odot}$ yr$^{-1}$) & (degeneracy)   \\
\hline 
e1s1 &  0.22 & 3.95$\pm$0.66 & 3.07$\pm$3.13 & 3.61$\pm$0.09 & 1.78$\pm$0.86 & 0.04$\pm$1.35  & -4.46$\pm$0.88 & -1.93$\pm$0.87 & -6.34$\pm$1.24 & 1265   \\   
e1s2 &  0.10 & 4.36$\pm$0.67 & 3.58$\pm$3.03 & 3.65$\pm$0.09 & 1.92$\pm$0.76 & 0.19$\pm$1.37  & -4.31$\pm$0.87 & -1.83$\pm$0.84 & -6.69$\pm$1.27 & 1628   \\  
e1s3 &  0.00 & 4.45$\pm$0.45 & 5.01$\pm$1.79 & 3.66$\pm$0.06 & 2.30$\pm$0.33 & 0.61$\pm$0.90  & -4.13$\pm$0.61 & -1.63$\pm$0.49 & -6.54$\pm$0.73 & 1396   \\ 
e1s4 &  0.02 & 4.79$\pm$1.04 & 4.52$\pm$2.50 & 3.76$\pm$0.22 & 2.25$\pm$0.54 &-0.98$\pm$2.86  & -4.50$\pm$1.93 & -1.90$\pm$0.90 & -6.80$\pm$1.27 & 2276   \\   
e2s1 &  0.03 & 3.99$\pm$0.84 & 4.54$\pm$3.90 & 3.70$\pm$0.22 & 2.33$\pm$0.80 & 0.13$\pm$1.82  & -4.30$\pm$1.31 & -1.82$\pm$0.87 & -5.98$\pm$1.34 & 1131   \\   
e2s2 &  0.01 & 4.40$\pm$1.11 & 2.24$\pm$2.18 & 3.68$\pm$0.19 & 1.54$\pm$0.76 &-1.19$\pm$2.28  & -4.90$\pm$1.56 & -2.28$\pm$0.84 & -6.92$\pm$1.25 & 2292    \\  
e2s3 &  0.04 & 4.22$\pm$0.84 & 7.37$\pm$4.38 & 3.85$\pm$0.32 & 2.98$\pm$0.83 & 1.04$\pm$1.66  & -3.95$\pm$1.06 & -1.73$\pm$1.03 & -5.94$\pm$1.40 &  771     \\ 
\hline 
\end{tabular}
\end{table*}

\begin{table*}
\centering
\caption{Mt. Abu J, H, K (MKO system) and {\it Spitzer}-IRAC Ch1, Ch2 and Ch3 band photometry (in mag) of the new 
Class I YSOs identified in the vicinity of S235A-B region using HK$_{s}$+Ch1-2 criteria (see text for details).}
\label{tab5}
\begin{tabular}{lccccccccccccccccr}
\hline
Object      &	RA [2000] &  Dec [2000] &  J & H &  K & Ch1& Ch2& Ch3   \\     
\hline    													     
      1     &  5:40:45.68 & 35:43:40.4 &16.78$\pm$0.08  & 16.08$\pm$0.05  & 15.21$\pm$ 0.1 &  13.89$\pm$0.01 &  13.34$\pm$0.01& 12.83$\pm$0.10   \\
      2     &  5:40:46.81 & 35:43:59.0 &16.97$\pm$0.08  & 15.93$\pm$0.05  & 15.63$\pm$0.13 &  15.04$\pm$0.03 &  14.32$\pm$0.02& 14.38$\pm$0.16   \\
      3     &  5:40:49.16 & 35:41:37.5 &	-	& 15.89$\pm$0.04  & 14.35$\pm$0.06 &  12.09$\pm$0.02 &  11.51$\pm$0.01& 10.48$\pm$0.02   \\
      4     &  5:40:52.23 & 35:41:17.3 &	-	& 15.28$\pm$0.04  & 14.31$\pm$0.04 &  12.49$\pm$0.05 &  11.76$\pm$0.03& 10.47$\pm$0.07   \\
      5     &  5:40:53.42 & 35:42:07.2 &13.66$\pm$0.03  & 12.28$\pm$0.04  & 11.05$\pm$0.04 &   9.07$\pm$0.04 &   8.57$\pm$0.03& 6.60 $\pm$0.08   \\
      6     &  5:40:53.69 & 35:41:47.5 &16.39$\pm$0.05  & 15.04$\pm$0.03  & 14.48$\pm$0.06 &  12.55$\pm$0.06 &  11.61$\pm$0.04& 	-        \\
      7     &  5:40:54.55 & 35:43:57.8 &16.35$\pm$0.07  & 14.61$\pm$0.07  & 13.56$\pm$0.06 &  12.36$\pm$0.04 &  11.65$\pm$0.04& 	-        \\
      8     &  5:40:55.07 & 35:43:59.0 &16.25$\pm$0.07  & 15.00$\pm$0.06  & 14.39$\pm$0.06 &  13.02$\pm$0.03 &  12.23$\pm$0.04& 	-        \\
      9     &  5:40:55.10 & 35:41:33.7 &	-	& 16.03$\pm$0.05  & 15.13$\pm$0.08 &  13.63$\pm$0.04 &  12.77$\pm$0.04& 12.15$\pm$0.19   \\
      10    &  5:40:55.94 & 35:43:07.8 &	-	& 16.17$\pm$0.06  & 15.44$\pm$0.11 &  13.92$\pm$0.05 &  13.48$\pm$0.04& 	-        \\
      11    &  5:40:56.54 & 35:42:39.5 &16.91$\pm$0.08  & 15.50$\pm$0.05  & 14.70$\pm$0.06 &  13.62$\pm$0.20 &  12.86$\pm$0.10& 	-        \\
\hline          
\end{tabular}
\end{table*}

\begin{table*}
\centering
\scriptsize
\caption{Mt. Abu J, H, K (MKO system) and {\it Spitzer}-IRAC Ch1, Ch2 and Ch3 band photometry (in mag) of the new Class II YSOs 
identified in the vicinity of S235A-B region using HK$_{s}$+Ch1-2 criteria (see text for details).}
\label{tab6}
\begin{tabular}{lccccccccccccccccr}
\hline
Object      &	RA [2000] &  Dec [2000] &  J & H &  K & Ch1& Ch2 & Ch3  \\     
\hline          											    	  	
           1  &  5:40:36.33 & 35:41:49.2 &	  -	   &	16.11$\pm$0.05   &   15.25$\pm$0.09  &    14.71$\pm$0.01  &    14.45$\pm$0.02 &   14.12$\pm$0.11  \\ 
           2  &  5:40:42.25 & 35:43:17.9 &	  -	   &	15.76$\pm$0.04   &   15.18$\pm$0.08  &    14.37$\pm$0.02  &    14.02$\pm$0.02 &   14.01$\pm$0.17  \\ 
           3  &  5:40:46.73 & 35:44:21.3 &16.14$\pm$0.04   &	15.21$\pm$0.03   &   14.78$\pm$0.07  &    14.37$\pm$0.02  &    14.22$\pm$0.02 & 	-	  \\ 
           4  &  5:40:47.03 & 35:42:23.2 &	  -	   &	14.95$\pm$0.04   &   13.36$\pm$0.03  &    11.61$\pm$0.02  &    11.04$\pm$0.01 &   10.71$\pm$0.12  \\ 
           5  &  5:40:47.59 & 35:42:21.8 &16.50$\pm$0.06   &	15.34$\pm$0.04   &   14.64$\pm$0.07  &    13.88$\pm$0.12  &    13.60$\pm$0.10 & 	-	  \\ 
           6  &  5:40:47.87 & 35:41:37.7 &	  -	   &	15.81$\pm$0.05   &   15.19$\pm$0.09  &    14.96$\pm$0.15  &    14.53$\pm$0.12 & 	-	  \\ 
           7  &  5:40:49.02 & 35:41:17.6 &	  -	   &	15.70$\pm$0.05   &   14.65$\pm$0.05  &    13.43$\pm$0.01  &    13.05$\pm$0.01 &   12.57$\pm$0.12  \\
\hline          												    
\end{tabular}													    
\end{table*}													    
														    
\begin{table*}													    
\centering													    
\scriptsize
\contcaption{}													    
\begin{tabular}{lccccccccccccccccr}
\hline
Object      &	RA [2000] &  Dec [2000] &  J & H &  K & Ch1& Ch2 & Ch3  \\     
\hline          											    	  	
          21  &  5:40:51.72 & 35:41:56.6 &	  -	   &	15.38$\pm$0.04   &   14.32$\pm$0.05  &    12.05$\pm$0.16  &    11.70$\pm$0.13 & 	-	  \\ 
          22  &  5:40:51.77 & 35:43:53.6 &	  -	   &	16.31$\pm$0.05   &   14.31$\pm$0.05  &    12.86$\pm$0.01  &    12.39$\pm$0.01 &   12.74$\pm$0.17  \\ 
          23  &  5:40:51.80 & 35:45:12.0 &	  -	   &	15.83$\pm$0.04   &   15.26$\pm$0.10  &    14.32$\pm$0.03  &    13.99$\pm$0.02 &   12.97$\pm$0.11  \\ 
          24  &  5:40:52.21 & 35:41:41.4 &15.18$\pm$0.03   &	14.06$\pm$0.05   &   13.17$\pm$0.05  &    11.78$\pm$0.05  &    11.34$\pm$0.04 & 	-	  \\ 
          25  &  5:40:52.34 & 35:42:42.1 &12.44$\pm$0.02   &	12.15$\pm$0.04   &   12.06$\pm$0.04  &    11.31$\pm$0.07  &    11.13$\pm$0.09 & 	-	  \\ 
          26  &  5:40:52.42 & 35:41:00.3 &15.98$\pm$0.04   &	15.08$\pm$0.04   &   14.43$\pm$0.06  &    13.75$\pm$0.03  &    13.58$\pm$0.05 & 	-	  \\ 
          27  &  5:40:52.66 & 35:41:45.4 &13.64$\pm$0.03   &	12.74$\pm$0.05   &   12.29$\pm$0.05  &    11.16$\pm$0.02  &    10.84$\pm$0.02 &   10.13$\pm$0.14  \\
\hline          
\end{tabular}
\end{table*}

\begin{table*}
\centering
\scriptsize
\caption{Some of the selected sources in S235A-B region; the numbers in parentheses in the Object column are common 
sources from Table~\ref{tab1} and ~\ref{tab2} designations (see Fig.~\ref{fig10} and text)}
\label{tab7}
\begin{tabular}{lcccccccccccccccccccr}
\hline
Object      &	RA [2000] &  Dec [2000] &  J & H &  K   &  Ch1    &   Ch2  &  Ch3   &  Ch4 &$\alpha_{IRAC}$ & Class\\
\hline       		   	                                      	      	      	 	 	 	   	   	     	    	     	      		 
s1          &  5:40:53.71 & 35:41:56.2 & 14.60$\pm$0.02  &13.18$\pm$0.04 & 12.38$\pm$0.04 &    11.01$\pm$0.04 &  11.22$\pm$0.16 &         -      &           -     &         -       &   -    	\\
s2(31)      &  5:40:53.49 & 35:41:52.6 & 16.21$\pm$0.07  &14.00$\pm$0.05 & 12.33$\pm$0.04 &    10.48$\pm$0.01 &   9.73$\pm$0.03 &   8.80$\pm$0.05&    7.66$\pm$0.06&	0.44$\pm$0.07&  0/I	\\
s3          &  5:40:53.16 & 35:41:59.3 &        -        &        -      &         -      &            -      &          -      &          -     &            -    &         -       &   -	 \\
s4          &  5:40:53.02 & 35:41:53.4 &        -        &        -      &         -      &            -      &  11.77$\pm$0.22 &          -     &            -    &         -       &   -	 \\
s5(MIR-AB)  &  5:40:53.44 & 35:41:48.8 &        -        &        -      &         -      &            -      &  11.39$\pm$0.05 &   9.66$\pm$0.08&    8.27$\pm$0.06&         -       &   -	 \\
s6          &  5:40:53.69 & 35:41:47.5 & 16.39$\pm$0.05  &15.04$\pm$0.03 & 14.48$\pm$0.06 &    12.56$\pm$0.06 &  11.61$\pm$0.04 &           -    &            -    &	     -       &   -	 \\   
s7(50;M1)   &  5:40:53.62 & 35:41:42.7 &        -        &15.28$\pm$0.05 & 13.46$\pm$0.04 &    11.25$\pm$0.01 &  10.16$\pm$0.01 &   9.55$\pm$0.04&    9.46$\pm$0.15&   -0.83$\pm$0.66&  II	\\
\hline 
\end{tabular}
\end{table*}

\begin{table*}
\centering
\scriptsize
\caption{Weighted mean values of the physical parameters derived from SED modeling of the labeled sources from Table~\ref{tab7} in S235A-B region (see text).
$\chi^{2}$ values are given per data point.}
\label{tab8}
\begin{tabular}{lcccccccccr}
    \hline 
Source  &$\chi^{2}$       & Age 	          & $M_*$	  &	$T_*$	    & $L_*$	      &  M$_{env}$	  &  $\dot{M}_{env}$		&  M$_{disk}$	    & $\dot{M}_{disk}$  	 &  No. of models \\
Name    &                & log(yr)	  & M$_{\odot}$   &log(T(K))	    & log(L$_{\odot}$)& log(M$_{\odot}$ ) & log(M$_{\odot}$ yr$^{-1}$)  & log(M$_{\odot}$ ) & log(M$_{\odot}$ yr$^{-1}$) &  (degeneracy) \\
\hline 
s1	  &    2.31  &  5.90$\pm$0.53 &   2.10$\pm$1.02 &    3.73$\pm$0.16 &   1.12$\pm$0.48 &  -3.40$\pm$2.15 & -6.32$\pm$1.52 &  -3.97$\pm$1.77 &  -9.91$\pm$1.88 &	6551	\\    
s2        &    0.44  &  6.16$\pm$0.76 &   3.27$\pm$1.22 &    3.93$\pm$0.20 &   1.89$\pm$0.42 &  -4.77$\pm$2.85 & -5.81$\pm$1.68 &  -3.64$\pm$1.79 &  -9.14$\pm$1.91 &	 737	\\   
s5        &    0.10  &  3.62$\pm$0.43 &  11.26$\pm$3.54 &    3.71$\pm$0.18 &   3.55$\pm$0.36 &   2.40$\pm$0.38 & -3.15$\pm$0.26 &  -0.93$\pm$0.77 &  -4.91$\pm$0.91 &	  62	 \\  
s6	  &    0.06  &  5.44$\pm$0.86 &   2.66$\pm$1.44 &    3.71$\pm$0.14 &   1.48$\pm$0.45 &  -1.76$\pm$2.51 & -4.74$\pm$2.24 &  -2.99$\pm$1.35 &  -8.42$\pm$1.62 &	 199	  \\ 
s7        &    1.45  &  5.21$\pm$0.43 &   2.40$\pm$1.06 &    3.66$\pm$0.07 &   1.48$\pm$0.24 &  -0.75$\pm$1.48 & -4.86$\pm$0.79 &  -2.22$\pm$0.85 &  -7.85$\pm$1.07 &	 364	\\   
s8        &    0.00  &  4.71$\pm$0.61 &   2.92$\pm$1.79 &    3.64$\pm$0.04 &   1.75$\pm$0.53 &   0.56$\pm$0.74 & -4.04$\pm$0.43 &  -2.05$\pm$0.83 &  -7.27$\pm$1.28 &	1849	\\   
S235B     &    5.10  &  4.59$\pm$0.54 &   6.51$\pm$1.34 &    3.76$\pm$0.13 &   2.76$\pm$0.35 &   0.95$\pm$0.61 & -3.81$\pm$0.53 &  -1.84$\pm$0.89 &  -6.66$\pm$1.14 &	  30	  \\ 
\hline 
\end{tabular}
\end{table*}

\end{document}